\documentclass[10pt]{article}
\usepackage{amsmath}
\usepackage{hyperref}

\usepackage[titletoc,title]{appendix}

\numberwithin{equation}{section}

\renewcommand{\a}{\alpha}
\newcommand{\w}{\omega}
\newcommand{\dd}{\mathrm{d}}

\newcommand{\pwo}{p_{\w_1}}
\newcommand{\pwt}{p_{\w_2}}

\begin{document}

\begin{flushright}
\vspace{1mm}
 FIAN/TD/18--2026\\
\vspace{-1mm}
\end{flushright}\vspace{1cm}

\begin{center}
{\large\bf Towards All-Order Locality in the Higher-Spin \\ Holomorphic Zero-Form Sector}

\vspace{1 cm}

{D.A.~Valerev\textsuperscript{1,2}}\\

\vspace{0.5 cm}
 \textsuperscript{1}I.E. Tamm Department of Theoretical Physics, Lebedev Physical Institute, \\ Leninsky prospect 53, 119991, Moscow, Russia

\vspace{1 cm}
 \textsuperscript{2}Moscow Institute of Physics and Technology, \\ Institutsky lane 9, 141700, Dolgoprudny, Moscow region, Russia\\

\end{center}

\begin{abstract}

This work addresses the construction of local interaction vertices in the holomorphic zero-form sector obtained from Vasiliev equations. We focus on a requirement on zero-form vertices, previously proposed in a less restrictive form in \cite{proj_comp}, which, together with spinor spin-locality, guarantees space-time spin-locality. Necessary and sufficient conditions on the holomorphic one-form sector vertices are derived that ensure the existence of holomorphic zero-form vertices satisfying this requirement. As a concrete demonstration, the proposed scheme is applied to compute the spin-local holomorphic $\eta^2$ zero-form sector vertex, a result that goes beyond previous work where only spinor spin-locality was established. These findings lay the groundwork for a systematic all-order construction of local vertices in higher-spin theory.

\end{abstract}

\newpage
\tableofcontents
\newpage

\section{Introduction}
\label{sec:Introduction}
\subsection{General background}
Higher-spin gauge theories represent a very natural and wishful extension of the gauge principles, which serve as the basis for building the Standard Model and general relativity. In particular, spin-1 gauge theory describes Maxwell electromagnetic fields, while spin-2 gauge fields represent the essence of gravity. Theory of free massless higher-spin fields was constructed by Fronsdal and Fang \cite{Fronsdal, Fang}. Nevertheless, the issue if there exists a consistent interacting theory of massless fields of spins $s>2$ took a decades of research. Such a theory, if it exists, should play a crucial role at higher-energy regimes \cite{wCC_local}, being a candidate for a role of quantum gravity.

A consistent theory on an AdS background was constructed in \cite{nonlinear_system}. One of the most intriguing questions about this system is its locality. However, as far as the theory contains an infinite number of fields, the standard definition of locality, i.e. that the number of derivatives in a vertex is bounded, fails to be adequate. It was found that the number of derivatives grows with the spin of the fields \cite{Bengtsson:1983pd}-\cite{Ders_grow_with_s}. Although the conventional definition has no chance of being fulfilled, one can formulate so-called \emph{spin-locality}, which means that a vertex for fields with fixed spins contains a finite number of derivatives. 

The question of locality has also been addressed from the holographic perspective. The AdS/CFT correspondence has been invoked to argue for the non-locality of higher-spin gauge theory. In particular, the Klebanov--Polyakov conjecture \cite{KlebanovPolyakov, SezginSundell}, which identifies the bulk higher-spin theory with the free 3d vector sigma model on the boundary, has been used to reconstruct bulk interactions from the boundary theory. These reconstructions led several authors to conclude that the higher-spin gauge theory must be essentially non-local beyond the leading order \cite{BekaertErdmengerPonomarevSleight}-\cite{Ponomarev}. However, for a number of reasons these results are questionable (not least because the conjecture itself remains unproved) and the issue of locality in the bulk theory is far from settled. In this respect, it is worth noting an alternative point of view \cite{Vasiliev_AdS_CFT}, according to which the boundary theory dual to higher-spin theory should not be a free sigma model, but rather a theory of 3d conformal currents of all spins interacting with 3d conformal higher-spin fields of Chern-Simons type. This approach is currently being developed in recent years, see e.g. \cite{Iazeolla_Sundell}. This underscores the importance of studying the locality problem directly within the bulk formulation, rather than relying solely on holographic arguments.

Motivated by this, a huge amount of work was undertaken to answer the question whether vertices are spin-local or "how much" they are nonlocal directly in the bulk. With the help of the developed formalism \cite{Homotopy_Operators}-\cite{diff_hom} for working with generating nonlinear higher-spin equations \cite{nonlinear_system}, several results on lower-order vertices were obtained \cite{wCC_local}, \cite{w2C}-\cite{ moderate_nonlocality}. However, it was not possible to obtain any results beyond the second order of perturbation theory with respect to the coupling constant\footnote{In the literature, the order of perturbation is also commonly counted by the number of fields present in a vertex rather than by powers of the coupling constant.} using the standard generating system \cite{nonlinear_system}. At the same time, for the holomorphic theory, stunning success was achieved with the help of the nonlinear generating system introduced by Didenko \cite{Didenko} (see also \cite{consistency}). It was proved that all vertices are local and, moreover, their explicit form was even found \cite{Povarnin}. Although this generating system currently produces only the vertices of the holomorphic higher-spin theory, ongoing research is actively directed towards extending it to a complete formulation \cite{irregular}. Beyond the perturbative construction of vertices, the generating system for the self-dual sector offers a setting in which the symmetry-breaking mechanism can be implemented, a mechanism that is thought to connect higher-spin theory to string theory. The system possesses exact solutions of high residual symmetry, whose linearised properties can be explored using the vertices obtained in \cite{Povarnin}; for details, we refer to \cite{Faliakhov_25, Faliakhov_26}. A distinct approach to the construction of local vertices in the self-dual (or chiral) theory was given in \cite{Sharapov_1, Sharapov_2}. 

All known spin-local holomorphic vertices, obtained from either the full nonlinear system or the holomorphic one, share several notable features. While some were studied in \cite{shift_symmetry}, the central role in this work belongs to the \emph{vertex dualities} of the one-form sector, identified in \cite{Povarnin}; these are in partial correspondence with the cyclic structure of the HS $A_\infty$-algebra in the sense of \cite{Sharapov_1}. We demonstrate here that these dualities, supplemented by certain additional assumptions, are both necessary and sufficient, in a meaningful sense, for the construction of spin-local zero-form vertices of a constrained form, to all orders. Whether spin-locality can also be achieved for vertices not of this form, i.e. without the corresponding one-form symmetries, remains an open question.

This paper is organized as follows. Section \ref{sec:Introduction} sets up the generating system and defines our notation. In section \ref{proj-comp_0-form_vertices}, we present a systematic procedure for obtaining spin-local vertices at all perturbative orders, subject to assumptions on the one-form sector. Section \ref{construction} then demonstrates how the given fixed consistent zero-form vertices can be reproduced within the generating system \cite{nonlinear_system}. We close in section \ref{sec:Conclusion} with a discussion of our results and an outlook on open problems.

\subsection{Generating system}
The Fronsdal equations for massless fields of spin~$s$ govern the dynamics of symmetric, double-traceless tensor fields $\phi^{a(s)}$ (we confine ourselves here to integer spins). Double tracelessness is defined by the condition
\begin{equation}
    \phi^{a_1 a_2 a_3 a_4 a(s-4)} g_{a_1 a_2} g_{a_3 a_4} = 0, \quad a_i = 0, \dots, 4,
\end{equation}
where $g_{ab}$ denotes the spacetime metric. The field equations are given by
\begin{equation}
    \partial_m \partial^m \phi^{a(s)} - \partial^{a} \partial_m \phi^{m a(s-1)} + \partial^a \partial^a \phi^{a(s-2) m}{}_m = 0.
    \label{Fronsdal_eqs}
\end{equation}
Following the notation of \cite{Vasiliev:1986qx}, in this section only, repeated indices at the same level (all lower or all upper) or indices accompanied by a bracketed number indicate full symmetrization; for example,
\begin{equation}
    \partial^a \phi^a := \frac{1}{2}\left(\partial^{a_1} \phi^{a_2} + \partial^{a_2} \phi^{a_1}\right); \quad \phi^{a(2)}  := \frac{1}{2}\left(\phi^{a_1 a_2} + \phi^{a_2 a_1}\right).
\end{equation}
However, the frame-like approach proves more convenient in many respects~\cite{Annals_89}. To describe all spins uniformly, one introduces formal variables $Y^A = (y^\alpha, \bar{y}^{\dot{\alpha}}), \quad \alpha, \dot{\alpha} = 1, 2,$
where $\alpha$ and $\dot{\alpha}$ are two-component spinor indices of opposite chirality. Along with these variables, one defines the Moyal star product
\begin{equation}
    (f * g)(Y) = \int dU \, dV \, f(Y+U) \, g(Y+V) \, \exp(i U_A V^A),
\end{equation}
with the integration measure
\begin{equation}
    dU = d^2 u \, d^2 \bar{u} = \frac{1}{2\pi} du^1 du^2 \, \frac{1}{2\pi} d\bar{u}^1 d\bar{u}^2,
\end{equation}
and similarly for $dV$ and all subsequent integrals. In this construction, one also introduces Klein operators $K = (k, \bar{k})$, which satisfy
\begin{equation}
\begin{gathered}
 \{k, y_\alpha\} = [k, \bar{y}_{\dot{\alpha}}] = [\bar{k}, y_\alpha] = \{\bar{k}, \bar{y}_{\dot{\alpha}}\} = 0,
 \\
k^2 = \bar{k}^2 = 1, \qquad [k, \bar{k}] = 0.
\end{gathered}
\label{Klein_Y}
\end{equation}
The $\mathrm{sp}(2)$ indices are raised and lowered using the symplectic structure:
\begin{equation}
A^\alpha = \epsilon^{\alpha \beta} A_\beta, \qquad 
A_\alpha = A^\beta \epsilon_{\beta \alpha}, \qquad 
\epsilon_{\alpha \beta} = -\epsilon_{\beta \alpha}, \qquad 
\epsilon^{\alpha \beta} = -\epsilon^{\beta \alpha}, \qquad 
\epsilon_{\alpha \beta} \epsilon^{\gamma \beta} = \delta_{\alpha}^{\gamma},     
\end{equation}
\begin{equation}
\bar{A}^{\dot{\alpha}} = \epsilon^{\dot{\alpha} \dot{\beta}} \bar{A}_{\dot{\beta}}, \qquad 
\bar{A}_{\dot{\alpha}} = \bar{A}^{\dot{\beta}} \epsilon_{\dot{\beta} \dot{\alpha}}, \qquad 
\epsilon_{\dot{\alpha} \dot{\beta}} = -\epsilon_{\dot{\beta} \dot{\alpha}}, \qquad 
\epsilon^{\dot{\alpha} \dot{\beta}} = -\epsilon^{\dot{\beta} \dot{\alpha}}, \qquad 
\epsilon_{\dot{\alpha} \dot{\beta}} \epsilon^{\dot{\gamma} \dot{\beta}} = \delta_{\dot{\alpha}}^{\dot{\gamma}},     
\end{equation}
where \(\epsilon_{\alpha \beta}\) and \(\epsilon_{\dot{\alpha} \dot{\beta}}\) denote the canonical symplectic metrics, given explicitly by
\begin{equation}
\epsilon_{\alpha \beta} = \begin{pmatrix} 0 & 1 \\ -1 & 0 \end{pmatrix}, \qquad
\epsilon_{\dot{\alpha} \dot{\beta}} = \begin{pmatrix} 0 & 1 \\ -1 & 0 \end{pmatrix}.     
\end{equation}
For notational economy, and whenever possible, we shall adopt the standard condensed convention for contracting two-component indices:
\begin{equation}
\xi_\alpha \eta^\alpha := \xi \eta = -\eta \xi, \qquad
\bar{\xi}_{\dot{\alpha}} \bar{\eta}^{\dot{\alpha}} := \bar{\xi} \bar{\eta} = -\bar{\eta} \bar{\xi}.     
\end{equation}
One then passes to generating functions in the variables $Y$,
\begin{equation}
     \omega(Y ,K|x) =  \sum_{j=0}^{1} \sum_{n=0,m=0}^{\infty} \frac{1}{2n!m!} \omega_{\alpha_1, \dots, \alpha_n, \dot{\alpha}_1, \dots, \dot{\alpha}_m}^{(j)}(x) 
    y^{\alpha_1} \cdots y^{\alpha_n} \bar{y}^{\dot{\alpha}_1} \cdots  \bar{y}^{\dot{\alpha}_m}  k^j \bar{k}^j,
\end{equation}
\begin{equation}
      C(Y ,K|x) =  \sum_{j=0}^{1} \sum_{n=0,m=0}^{\infty} \frac{1}{2n!m!} C_{\alpha_1, \dots, \alpha_n, \dot{\alpha}_1, \dots, \dot{\alpha}_m}^{(j,1-j)}(x) 
    y^{\alpha_1} \cdots y^{\alpha_n} \bar{y}^{\dot{\alpha}_1} \cdots  \bar{y}^{\dot{\alpha}_m}  k^j \bar{k}^{1-j},
\end{equation}
which describe, respectively, the one-form $\omega$, containing all gauge fields of spin $\geq 1$ and their derivatives, and the zero-form $C$, containing the HS Weyl tensors, lower-spin matter fields, and all associated derivatives. The Fronsdal equations can then be equivalently reformulated \cite{Annals_89} in terms of the introduced generating functions $\omega$ and $C$ as
\begin{equation}
    \begin{aligned}
        \mathrm{d}_x \omega + \left\{ \omega, \Omega \right\}_* &= \Upsilon(\Omega, \Omega, C), 
        \\
        \mathrm{d}_x C + \left[\Omega, C \right]_* &= 0
    \end{aligned}
\end{equation}
for a specific $\Upsilon(\Omega, \Omega, C)$, where $\Omega$ denotes the AdS background connection. These equations constitute the essence of the central on-mass-shell theorem (see \cite{Vasiliev:1999, Elements} for details). They can be interpreted as the linearization near the vacuum solution of empty AdS space, $\omega = \Omega, \; C = 0$, of the full unfolded nonlinear higher-spin equations
\begin{align}
    \dd_x \w + \w*\w =& \Upsilon(\w^2 C) + \Upsilon(\w^2 C^2) + \dots, \\
    \dd_x C + \left[\w, C \right]_* =& \Upsilon(\w C^2) + \Upsilon(\w C^3) + \dots.
\end{align}
A natural question that arises is how these nonlinear vertices $\Upsilon$ can be constructed in a consistent manner. To address this issue, the nonlinear generating higher-spin equations were formulated in \cite{nonlinear_system} as the system
\begin{align}
       \dd_x &W + W * W = 0,
    \label{master_eqs_1} \\
      \dd_x &B + \left[ W, B\right]_* = 0,
    \label{master_eqs_2} \\
     -2i \dd_Z &W +    \dd_x S + \left\{W ,S\right\}_* = 0,
     \label{master_eqs_3} \\
    -2i \dd_Z &B + \left[S, B\right]_* = 0,
    \label{master_eqs_4} \\
      -2i \dd_Z& S + S *S = i B * (\eta \gamma + \bar{\eta} \bar{\gamma}).
    \label{master_eqs_5}
\end{align}
The fields $W(Z, Y, K|x)$, $S(Z, Y, K|x)$, and $B(Z, Y, K|x)$ depend on several sets of variables. First, they depend on the spacetime coordinates $x^{\underline{n}}$, equipped with the standard de Rham differential $\mathrm{d}_x := dx^{\underline{n}} \frac{\partial}{\partial x^{\underline{n}}}$. Second, they depend on the aforementioned spinor variables $Y^A = (y^\alpha, \bar{y}^{\dot{\alpha}})$; no differentials are taken with respect to these variables. Third, they depend on another set of auxiliary spinor variables $Z^A = (z^\alpha, \bar{z}^{\dot{\alpha}})$, with the associated differential $\mathrm{d}_Z = dZ^A \frac{\partial}{\partial Z^A} =: \theta^A \frac{\partial}{\partial Z^A}$. Finally, they depend on the Klein operators $(k, \bar{k})$, which satisfy in addition to \eqref{Klein_Y} the commutation relations
\begin{equation}
\{k, z_\alpha\} = [k, \bar{z}_{\dot{\alpha}}] = [\bar{k}, z_\alpha] = \{\bar{k}, \bar{z}_{\dot{\alpha}}\} = 0.
\end{equation}
The generating fields carry different differential-form gradings: $W$ is a 1-form in $x$ and a 0-form in $Z$, $S$ is a 0-form in $x$ and a 1-form in $Z$, while $B$ is a 0-form in both $x$ and $Z$.

The star product of two functions $f(Z; Y), g(Z; Y)$ is generalized as
\begin{equation}
    (f*g)(Z; Y) = \int dU dV f(Z+U; Y+U) g(Z-V;Y+V) \exp i U_A V^A.
    \label{star_prd_1}
\end{equation}
One can also separate the product into unbarred and barred parts:
\begin{align}
    (f*g)(z, \bar{z}; y, \bar{y}) &=  \int du dv f(z+u, \bar{z}; y+u, \bar{y}) \; \bar{*} \;  g(z-v, \bar{z};y+v, \bar{y}) \exp i u v,
    \\
    (f \; \bar{*} \; g)(z, \bar{z}; y, \bar{y}) &=  \int d\bar{u} d\bar{v} f(z, \bar{z} + \bar{u}; y, \bar{y}+\bar{u})  g(z, \bar{z} - \bar{v};y, \bar{y} + \bar{v}) \exp i \bar{u} \bar{v}.
    \label{star_prd_3}
\end{align}
The algebra of formal power series (naively setting aside the question of functional classes) in $Z,Y$ with the star product possesses central elements
\begin{equation}
     \gamma = \exp(i z y) k \theta^\alpha \theta_\alpha, \quad
     \bar{\gamma} = \exp(i \bar{z} \bar{y}) \bar{k} \bar{\theta}^{\dot{\alpha}} \bar{\theta}_{\dot{\alpha}}.
\end{equation}
That is, $\gamma * f = f * \gamma$ and $\bar{\gamma} * f = f * \bar{\gamma}$ for all $f$.

The procedure for obtaining interaction vertices is remarkably straightforward. One proceeds by expanding all fields perturbatively in powers of the coupling constant $\eta$:
\begin{align}
    W(Z,Y) &= \omega(Y) + W_1(Z,Y) + \dots,
    \\
    B(Z,Y) &= C(Y) + B_2(Z,Y) + \dots,
    \\
    S(Z,Y) &= 0 + S_1(Z,Y) +  \dots.
\end{align}
Here we denote the $Z$-independent components $W_0$ and $B_1$ by $\omega$ and $C$, respectively. As we shall demonstrate momentarily, these components are indeed independent of the $Z$-variables. A word of caution is in order regarding a potential source of confusion arising from historical conventions. Specifically, the perturbative orders scale as $W_n \propto |\eta|^n$ and $S_n \propto |\eta|^n$, whereas $B_n \propto |\eta|^{n-1}$. With the vacuum solution chosen such that $S_0 = 0$, it follows from \eqref{master_eqs_3} and \eqref{master_eqs_4} that $\mathrm{d}_Z W_0 = \mathrm{d}_Z B_1 = 0$. It is straightforward to verify that, to all orders, equations \eqref{master_eqs_3}--\eqref{master_eqs_5} reduce to the following system to be solved:
\begin{equation}
\begin{gathered}
        -2i\,\mathrm{d}_Z S_n = i B_n * (\eta \gamma + \bar{\eta} \bar{\gamma}) - \sum_{k=1}^{n-1} \left\{S_k, S_{n-k} \right\}_* ; 
        \qquad 
        2i\,\mathrm{d}_Z W_n =  \sum_{k=0}^{n-1} \left\{W_k, S_{n-k} \right\}_* + \sum_{k=1}^n \mathrm{d}_x S_n \big|^{\omega C^n};
        \\
        2i\,\mathrm{d}_Z B_n =  \sum_{k=1}^{n-1} \left[S_k, B_{n-k} \right]_*.
\end{gathered}
\label{master-field_eqs}
\end{equation}
The desired interaction vertices are then obtained by substituting the solutions for $W_n$ and $B_n$ into \eqref{master_eqs_1} and \eqref{master_eqs_2}:
\begin{equation}
    \Upsilon(\omega^2 C^n) = -\sum_{k=0}^n \left\{ W_k, W_{n-k} \right\}_* - \sum_{k=1}^n \mathrm{d}_x W_k \bigg|^{\omega^2 C^n}; 
    \qquad 
     \Upsilon(\omega C^n) =  -\sum_{k=0}^{n-1} \left[W_k, B_{n-k} \right]_* - \sum_{k=2}^n \mathrm{d}_x B_k \bigg|^{\omega C^n}.
     \label{vertices}
\end{equation}
Although the master-fields $W_n$ and $B_n$ depend on the $Z$-variables, the system \eqref{master_eqs_1}--\eqref{master_eqs_5} guarantees that the resulting vertices \eqref{vertices} are $Z$-independent. For further analysis of \eqref{master-field_eqs} and \eqref{vertices}, see \cite{Sezgin:2002ru, Sezgin:1998eh}. 
\subsection{Integral kernels notation}
All fields we face during resolving the holomorphic sector\footnote{This sector was identified in \cite{nonlinear_system} as the self-dual sector.} (the vertices proportional solely to $\eta^n$ and independent of $\bar{\eta}$) of the generating system can be represented with the help of integral kernels with respect to holomorphic variables: 
\begin{equation}
    B_n(z,y) = \int \left( \prod_j d^2p_j d^2r_j e^{-i p_j r_j} \right)  B_n(z, y|p_1, \dots, p_n) C(r_1) \bar{*} C(r_2) \bar{*} \dots \bar{*} C(r_n
    ),
\end{equation}
\begin{equation}
    \delta B_n(y) = \int \left( \prod_j d^2p_j d^2r_j e^{-i p_j r_j} \right)  \delta B_n(y|p_1, \dots, p_n) C(r_1) \bar{*} \dots \bar{*} C(r_n),
    \label{deltaB_n_kernel_notation}
\end{equation}
\begin{multline}
    \Upsilon_n^{[k]}(y) =  \int \left( \prod_j d^2p_j d^2r_j e^{-i p_j r_j} \right) \Upsilon_n^{[k]} (y|p_\w;p_1, \dots, p_n) \times
    \\
    \times  C(r_1) \bar{*} \dots \bar{*} C(r_{k}) \bar{*} \w(r_\w) \bar{*} C(r_{k+1}) \bar{*} \dots \bar{*} C(r_n) k^{n+1},
\end{multline}
\begin{multline}
    \Upsilon_n^{[k_1, k_2]}(y) =  \int \left( \prod_j d^2p_j d^2r_j e^{-i p_j r_j} \right) \Upsilon_n^{[k_1, k_2]}(y|p_{\w_1}, p_{\w_2};p_1, \dots, p_n) \times 
    \\
    \times C(r_1) \bar{*} \dots  \bar{*} C(r_{k_1}) \bar{*} \w(r_{\w_1})\bar{*} C(r_{k_1 + 1}) \bar{*} \dots \bar{*} C(r_{k_2}) \bar{*} \w(r_{\w_2}) \bar{*} C(r_{k_2 + 1}) \bar{*} \dots C(r_{n}) k^{n}.
\end{multline}
We will denote the field and its integral kernel by the same symbol: there is a trivial one-to-one correspondence between them, so this will not lead to any ambiguities. $\delta B_n$ \eqref{deltaB_n_kernel_notation} is mentioned here to demonstrate the notation for $Z$-independent functions. We also note that each variable $p_k$ is, up to a coefficient, the Fourier image of the derivative with respect to the field, whose argument has been replaced by $r_k$. We emphasize that, whereas the dependence of the fields $\omega$ and $C$ on $\bar{y}$ and $K = (k, \bar{k})$  is left implicit, the integral kernels have a restricted $K$-dependence, as we are concerned only with dynamical fields. For a recent analysis of the topological field sector, the reader is referred to \cite{Kirakosiants}.

Moreover, in practice, the integral kernels have the form
\begin{equation}
 F(z,y|p_{\w_a}; p_j) = \int d\tau \mu(\tau) P(z,y|p_{\w_a}; p_j |\tau) \cdot e^{i \left[ T(\tau) zy + \sum_j \left(A^j(\tau) z + B^j(\tau) y  \right)p_j  +  P^{ij}(\tau) p_i p_j\right]}, \quad i,j = \w_1, \w_2, \dots, 1, 2\dots,
 \label{preexp_exp_form}
\end{equation}
where $P$ is a polynomial of finite degree in its arguments, and $\tau$ denotes a set of integration variables integrated against a compactly supported measure $\mu$.

It is appropriate here to discuss various notions of locality, following \cite{1910}. We call an expression \emph{spinor spin-local} if, in an expression of the form \eqref{preexp_exp_form}, the exponential contains no $\mathrm{sp}(2)$-convolutions of derivatives with respect to the fields $C$: $p_i p_j$, $i,j=1, \dots, n$. This means that, for fixed spins of the fields $\omega$ and $C$, the expression contains only a finite number of derivatives with respect to auxiliary arguments such as $y$ of these fields. This is easy to see, since for a fixed spin $\omega(y)$ is a polynomial of finite degree, while $C(y)$ is an infinite power series in the variables $y$ (for details, cf. \cite{Vasiliev:1999}). Moreover, this definition should not be confused with the \emph{space-time spin-locality} (further noted just as \emph{spin-locality}), i.e., the property that, upon fixing the spins of the fields, the expression contains a finite number of space-time derivatives. Of course, in a physical context, it is precisely space-time spin-locality that is important to us; however, in practice, it is more convenient to control spinor spin-locality. How these two notions are related is a good question, discussed in \cite{proj_comp}.

An expression is called \emph{ultralocal} \cite{w2C} if, in its exponential, in addition to the convolutions of derivatives with respect to the fields $C$, there are also no convolutions of derivatives with respect to the fields $C$ with the variable $y$, i.e., $y p_i$, $i=1, \dots, n$. This property is more restrictive than spinor spin-locality, but it often arises in the study of vertices and is convenient for analysis.

\section{Projectively-compact holomorphic zero-form vertices}
\label{proj-comp_0-form_vertices}
 In this section, we provide an algorithm for obtaining spin-local vertices of the holomorphic one-form sector to all orders in perturbation theory. As shown in \cite{proj_comp}, spinor spin-locality of vertices does not guarantee the desired space-time spin-locality. However, several cases have been found where spinor spin-locality does imply space-time spin-locality. Restricting our attention to the holomorphic zero-form sector vertices, we wish to discuss a particularly interesting case. Although it is not the most general condition one could impose, it has turned out to be very effective and, precisely because of its restrictiveness, remarkably efficient to verify. This condition has the form
\begin{equation}
\Upsilon(\omega C^n) \big|_{y=0} = 0.
\label{U_(y=0)=0}
\end{equation}
Notably, all holomorphic zero-form sector vertices found in \cite{Povarnin} fulfill \eqref{U_(y=0)=0}.
Thus, our goal is to obtain spinor spin-local holomorphic zero-form sector vertices that satisfy \eqref{U_(y=0)=0}. 

 Let us consider, by induction, the $n$-th order of perturbations theory in powers of the $C$-field. This means that the auxiliary fields $S_1, \dots, S_{n-1}$, $W_0 = \w(y), W_1, \dots, W_{n-1}$, and $B_1 = C(y), B_2, \dots, B_{n-1}$ are found and fixed. Moreover, these fields give rise to spinor spin-local lower-order vertices $\Upsilon(\w C^2), \dots, \Upsilon(\w C^{n-1})$ that satisfy \eqref{U_(y=0)=0}. We now consider the vertex \eqref{vertices}
 \begin{equation}
     \Upsilon(\w C^n) = - \sum_{j=2}^{n} \dd_x B_j \big|^{\w C^n} - \sum_{j=0}^{n-1}\left[W_j, B_{n-j} \right]_*.
 \end{equation}
 As it can be easily seen, the only additional  field we need to determine in order to calculate this vertex is $B_n$. The equation for $B_n$ reads \eqref{master-field_eqs}
 \begin{equation}
     2i \dd_zB_n = \sum_{j=1}^{n-1} \left[S_j, B_{n-j} \right]_*.
     \label{Bn_equation}
 \end{equation}
 Let us denote by $\overset{\circ}{B_n}$ any solution of equation \eqref{Bn_equation}. Then, since the zero de-Rham cohomology consists of constants, the general solution of \eqref{Bn_equation} can be represented as 
 \begin{equation}
     B_n(z,y) = \overset{\circ}{B_n} (z,y) + \delta B_n (y),
 \end{equation}
 where $\delta B_n(y)$ is an arbitrary $z$-independent function. If, by choosing $\overset{\circ}{B_n}$ as a particular solution, we obtain vertex $\overset{\circ}{\Upsilon}(\w C^n)$, then shift $B_n(z,y) = \overset{\circ}{B_n} (z,y) + \delta B_n (y)$ results in
 \begin{equation}
     \Upsilon(\w C^n) = \overset{\circ}{\Upsilon}(\w C^n) - \dd_x \delta B_n \big|^{\w C^n} -\left[\w, \delta B_n \right]_*.
 \end{equation}
In this section, we find shift $\delta B_n$ such that
\begin{equation}
    \left(\dd_x \delta B_n \big|^{\w C^n} + \left[\w, \delta B_n \right]_* \right)\bigg|_{y=0} =  \overset{\circ}{\Upsilon}(\w C^n) \big|_{y=0},
    \label{y=0_equation}
\end{equation}
which, in the perturbation order under consideration, implies \eqref{U_(y=0)=0}. However, such a shift exists only if the one-form sector vertices satisfy certain requirements (the issue of the necessity of these conditions is discussed in Appendix \ref{necessity}).

\subsection{Requirements for the one-form sector vertices}
 We impose several requirements on the one-form sector vertices. These requirements can be divided into two groups. The first group describes symmetry properties of the one-form sector vertices under cyclic permutations of $y$ and the derivatives of the $C$-fields (usually denoted, up to a factor, as $p_i$). Specifically, for all $0<k_1 \leq k_2 < n$, we demand that the following equalities hold 
\begin{align}
    \Upsilon_{n-1}^{[k_1, k_2]}((-1)^n p_n| \pwo, \pwt; p_1, \dots, p_{n-1}) = \Upsilon_{n-1}^{[k_1 - 1, k_2 - 1]}(p_1| \pwo, \pwt; p_2, \dots, p_n), 
    \label{requirement_cyclic} \\
    \Upsilon_{n-1}^{[0, k_1]}((-1)^n p_n| (-1)^n \pwt, \pwo; p_1, \dots, p_{n-1}) = -\Upsilon_{n-1}^{[k_1 - 1, n-1]}(p_1| \pwo, \pwt; p_2, \dots, p_n).
    \label{requirement_cyclic_2}
\end{align}
The second group, in turn, describes the prefactor of the one-form sector vertices. Namely, 
\begin{equation}
    \Upsilon_{n-1}^{[k_1, k_2]}(y|0,0; p_1, \dots,p_{n-1}) = 0.
    \label{requirement_nulling} 
\end{equation}
Though these assumptions could seem poorly motivated, there are three reasons why imposing these requirements is justified. Firstly, the lower-order vertices $\Upsilon(\omega^2 C)$ and $\Upsilon(\omega^2 C^2)$, found in \cite{w2C} and \cite{1909}, satisfy all of these constraints. Secondly, the holomorphic one-form sector vertices at all orders, computed in \cite{Povarnin}, also satisfy these requirements. Nevertheless, whether the vertices from \cite{Povarnin} are restrictions to the holomorphic sector of the full nonlinear unfolded higher-spin vertices (including the mixed sector) remains an open question. Finally, as shown in Appendix \ref{necessity}, if these conditions do not hold, then there is no way to obtain zero-form vertices with the target property \eqref{U_(y=0)=0}.

\subsection{Generalized vertex duality}
Let us consider nonlinear unfolded higher-spin equations \cite{Annals_89}
\begin{align}
    \dd_x \w + \w*\w =& \Upsilon(\w^2 C) + \Upsilon(\w^2 C^2) + \dots, \label{FrameLike_1} \\
    \dd_x C + \left[\w, C \right]_* =& \Upsilon(\w C^2) + \Upsilon(\w C^3) + \dots. \label{FrameLike_2}
\end{align}
Following the procedure of \cite{Povarnin}, we apply the differential $\dd_x$ to \eqref{FrameLike_2} and then set $y=0$.  Using our assumptions
\begin{equation}
    \Upsilon_m^{[k]} \big|_{y=0} = 0 \quad \forall m<n   
    \label{lower_order_vanishing_assumption}
\end{equation}
we obtain the resulting equalities in different orderings. We shall refer to these expressions as \emph{generalized vertex dualities}, given by equations \eqref{wwC^.}-\eqref{C^.wC^.w}, which with vanishing RHSs reproduce the ordinary \emph{vertex dualities} of \cite{Povarnin} obtained by setting $\Upsilon_n^{[k]} \big|_{y=0} = 0$.

The nonlinearities in \eqref{FrameLike_1} and \eqref{FrameLike_2} are governed by the $A_\infty$ structure of the higher-spin algebra \cite{Annals_89}. Since the $*$-product is associative and the vertices $\Upsilon$ organize into $A_\infty$ multilinear products, the Stasheff identities \cite{Stasheff1963I,Stasheff1963II} guarantee the consistency of the unfolded system order by order. The fields $\w$ and $C$ can therefore be regarded as taking values in an arbitrary associative algebra \cite{Annals_89}. This justifies treating each ordering $C^j \w C^{k-j} \w C^{n-k}$ separately in the derivation of the generalized vertex dualities below.
\vspace{0.5 cm}
\\
\noindent \textbf{Ordering} $\w \w C^n$:
\\
After applying $\dd_x$ to \eqref{FrameLike_2}, we obtain, in the $\w \w C^n$ ordering,
\begin{equation}
    \left(\dd_x \w \right) \big|^{\w \w C^{n-1}} * C - \w * \left( \dd_x C \right)\big|^{\w C^n} = \sum_{m} \left(\dd_x \Upsilon_m^{[0]} \right) \big|^{\w \w C^n}.
\end{equation}
Since setting $y=0$ commutes with $\dd_x$, and taking into account \eqref{lower_order_vanishing_assumption}, we arrive at
\begin{equation}
    \left(\Upsilon_{n-1}^{[0,0]} * C \right)\big|_{y=0} - \left(\w * \Upsilon_n^{[0]} \right)\big|_{y=0} =- \Upsilon_n^{[0]}(\w*\w, C, C, \dots)\big|_{y=0} + \Upsilon_n^{[0]}(\w, \w *C, C, \dots) \big|_{y=0},
\end{equation}
where we used that $\Upsilon_n^{[k]}$ depends on the space-time coordinates only via $\w(Y,K|x)$ and $C(Y,K|x)$. With the help of the formulas from Appendix \ref{useful_formulas}, this takes the following form in terms of integral kernels:
\begin{multline}
    \Upsilon_{n-1}^{[0, 0]} ((-1)^n p_n|\pwo,  \pwt;p_1, \dots, p_{n-1}) - \Upsilon_n^{[0]} (\pwo| \pwt; p_1, \dots, p_n) = 
    \\
    = - \Upsilon_n^{[0]} (0|\pwo + \pwt ; p_1, \dots, p_n) e^{-i \pwo \pwt} + \Upsilon_n^{[0]}(0|\pwo; \pwt + p_1, p_2, \dots, p_n) e^{-i \pwt p_1}.
    \label{wwC^.}
\end{multline}
Equations in all other orderings can be obtained in the same manner, therefore we omit the details of the calculations and present only the results:
\\
\noindent \textbf{Ordering} $\w C^k \w C^{n-k}, \; 0< k<n $:
\begin{multline}
    \Upsilon_{n-1}^{[0, k]} ((-1)^n p_n|\pwo,  \pwt;p_1, \dots, p_{n-1}) - \Upsilon_n^{[k]} (\pwo| \pwt; p_1, \dots, p_n) = 
    \\
    =- \Upsilon_n^{[0]} (0|\pwo; p_1, \dots, p_k + \pwt, \dots, p_n) e^{-i p_k \pwt} + \Upsilon_n^{[0]}(0|\pwo; p_1, \dots, \pwt + p_{k+1}, \dots,  p_n) e^{-i \pwt p_{k+1}} - 
    \\
     -\Upsilon_n^{[k]}(0|\pwt; \pwo + p_1, \dots, p_n) e^{-i \pwo p_1};
     \label{wC^.wC^.}
\end{multline}
\noindent \textbf{Ordering} $C^k \w \w C^{n-k}, \; 0< k<n$:
\begin{multline}
    \Upsilon_{n-1}^{[k, k]}((-1)^n p_n| \pwo, \pwt; p_1, \dots, p_{n-1}) - \Upsilon_{n-1}^{[k-1, k-1]}(p_1|\pwo, \pwt; p_2, \dots, p_n) = 
    \\
    = \Upsilon_n^{[k]}(0|\pwt; p_1. \dots, p_k + \pwo, p_{k+1}, \dots, p_n)e^{-i p_k \pwo} - \Upsilon_n^{[k]}(0|\pwo + \pwt; p_1, \dots, p_n) e^{-i \pwo \pwt} +
    \\
    +\Upsilon_n^{[k]}(0|\pwo; p_1. \dots, p_k , \pwt+  p_{k+1}, \dots, p_n)e^{-i \pwt p_{k+1}};
    \label{C^.wwC^.}
\end{multline}
\noindent \textbf{Ordering} $C^{j}\w C^{k-j} \w C^{n- k}$, \quad $0< j < k <n$:
\begin{multline}
    \Upsilon_{n-1}^{[j,k]}((-1)^n p_n| \pwo, \pwt; p_1, \dots, p_{n-1}) - \Upsilon_{n-1}^{[j - 1,k - 1]}(p_1| \pwo, \pwt; p_2, \dots, p_n) = 
    \\
    =\Upsilon_n^{[k]}(0|\pwt; p_1, \dots, p_{j} + \pwo, \dots, p_n) e^{-i p_{j} \pwo} - \Upsilon_n^{[k]}(0|\pwt; p_1, \dots,  \pwo + p_{j + 1}, \dots, p_n) e^{-i \pwo p_{j + 1}} -
    \\
    - \Upsilon_n^{[j]}(0|\pwo; p_1, \dots, p_{k} + \pwt, \dots, p_n) e^{-i p_{k} \pwt} + \Upsilon_n^{[j]}(0|\pwo; p_1, \dots, \pwt + p_{k + 1}, \dots, p_n) e^{-i \pwt p_{k + 1}};
    \label{C^.wC^.wC^.}
\end{multline}
\noindent \textbf{Ordering} $C^n \w \w$:
\begin{multline}
    -\Upsilon_{n-1}^{[n-1, n-1]}(p_1|\pwo, \pwt; p_2, \dots, p_{n-1}) - \Upsilon_n^{[n]}((-1)^n \, \pwt| \pwo; p_1, \dots, p_n) = 
    \\
    = \Upsilon_n^{[n]} (0| \pwt; p_1, \dots, p_n + \pwo) e^{-i p_n \pwo} - \Upsilon_n^{[n]}(0|\pwo + \pwt; p_1, \dots, p_n) e^{-i \pwo \pwt};
    \label{C^.ww}
\end{multline}
\noindent \textbf{Ordering} $\w C^n \w$:
\begin{multline}
    -\Upsilon_n^{[n]}(\pwo|\pwt; p_1, \dots, p_n) - \Upsilon_n^{[0]} ( (-1)^n \pwt| \pwo; p_1, \dots, p_n) =
    \\
    =- \Upsilon_n^{[0]}(0|\pwo; p_1, \dots, p_n + \pwt) e^{-i p_n \pwt} - \Upsilon_n^{[n]}(0|\pwt; \pwo + p_1, \dots, p_n) e^{-i \pwo p_1};
    \label{wC^.w}
\end{multline}
\noindent \textbf{Ordering} $C^k \w C^{n-k} \w, \; 0< k<n$:
\begin{multline}
     -\Upsilon_{n-1}^{[k-1, n-1]}(p_1|\pwo, \pwt; p_2, \dots , p_n) - \Upsilon_n^{[k]}((-1)^n \pwt| \pwo; p_1, \dots ,p_n) =
      \\
    = \Upsilon_n^{[n]} (0|\pwt; p_1, \dots, p_k + \pwo, p_{k+1}, \dots, p_n) e^{-i p_k \pwo} -  \Upsilon_n^{[n]} (0|\pwt; p_1, \dots, p_k , \pwo + p_{k+1}, \dots, p_n) e^{-i  \pwo p_k } +
    \\
    +\Upsilon_n^{[k]}(0|\pwo; p_1, \dots, p_n + \pwt) e^{-i p_n \pwt}.
    \label{C^.wC^.w}
\end{multline}
\subsection{First ordering: \texorpdfstring{$\w C^n$}{wCn}}
As was stated before, we consider a particular solution $\overset{\circ}{B_n}$ which leads to the vertex $\overset{\circ}{\Upsilon}(\w C^n)$. Let us check, that a shift 
\begin{equation}
    \delta B_n = \sum_{k=0}^{n-1} \overset{\circ}{\Upsilon} {}_n^{[k]}(0| y + \pi_{k}; 0, \dots, 0, p_{k+1}, \dots, p_n) e^{-i \left(y \pi_k + \Sigma_k \right)},
    \label{Bn_shift}
\end{equation}
where
\begin{equation}
    \pi_k = \sum_{j=1}^k p_j, \qquad \Sigma_k = \sum_{i<j}^k p_i p_j,
    \label{p_notations}
\end{equation}
resolves equation \eqref{y=0_equation} in the first $\w C^n$ ordering. Indeed,
\begin{equation}
\begin{aligned}
    \big(\dd_x &\delta B_n \big|^{\w C^n}  + \w * \delta B_n \big) \bigg|_{y=0} =
    \\
    &=\delta B_n(p_\w| p_1, \dots, p_n) - \delta B_n (0|p_\w + p_1, p_2, \dots, p_n) e^{-i p_\w p_1} = 
    \\
    &=\overset{\circ}{\Upsilon} {}_n^{[0]}(0| p_\w; p_1, \dots, p_n) + \sum_{k=1}^{n-1} \overset{\circ}{\Upsilon} {}_n^{[k]}(0| p_\w + \pi_{k}; 0, \dots, 0, p_{k+1}, \dots, p_n) e^{-i \left(p_\w \pi_k + \Sigma_k \right)} - 
    \\
    &- \overset{\circ}{\Upsilon} {}_n^{[0]}(0| 0; p_\w + p_1 , \dots, p_n) e^{-i p_\w p_1} - \sum_{k=1}^{n-1} \overset{\circ}{\Upsilon} {}_n^{[k]}(0| p_\w + \pi_{k}; 0, \dots, 0, p_{k+1}, \dots, p_n) e^{-i \left(p_\w \pi_k + \Sigma_k \right)} = 
    \\
    &\phantom{- \overset{\circ}{\Upsilon} {}_n^{[0]}(0| 0; p_\w + p_1 , \dots, p_n) e^{-i p_\w p_1}}=\overset{\circ}{\Upsilon} {}_n^{[0]}(0| p_\w; p_1, \dots, p_n) - \overset{\circ}{\Upsilon} {}_n^{[0]}(0| 0; p_\w + p_1 , \dots, p_n) e^{-i p_\w p_1}.
\end{aligned}
\end{equation}
But setting $\pwo = \pwt = 0$ in the equation \eqref{wwC^.} makes two summands on the RHS cancel each other, leading to
\begin{equation}
    \Upsilon_n^{[0]} (0| 0; p_1, \dots, p_n) = \Upsilon_{n-1}^{[0, 0]} ((-1)^n p_n|0,  0; p_1, \dots, p_{n-1}) = 0, 
\end{equation}
where the last equality holds due to the requirement \eqref{requirement_nulling}. So, it proves \eqref{y=0_equation} in the first $\w C^n$ ordering.

\subsection{Intermediate orderings: \texorpdfstring{$C^k \w C^{n-k}, \; 0< k< n$}{CkwC(n-k), 0<k<n}}
Proceeding further, let us show that the same shift \eqref{Bn_shift} resolves equation \eqref{y=0_equation} for all intermediate orderings $C^k \w C^{n-k}, \; 0< k< n$. Notice, that summand $\left[\w, \delta B_n \right]_*$ does not contribute to all of these ordering, since it yields only terms of the form $\w C^n$ and $C^n \w$. So, for any $0<k<n$
\begin{multline}
\begin{aligned}
    \dd_x \delta & B_n \big|_{y=0}^{C^k\w C^{n-k}} =
    \\
    &= 
    \delta B_n(0|p_1, \dots, p_k + p_\w, p_{k+1}, \dots, p_n) e^{-i p_k p_\w} - \delta B_n(0|p_1, \dots, p_k , p_\w + p_{k+1}, \dots, p_n) e^{-i p_\w p_{k+1}} =  \\
    &=\sum_{j < k} \overset{\circ}{\Upsilon}{}_n^{[j]}(0| \pi_j; 0, \dots, 0, p_{j+1}, \dots, p_k + p_\w, p_{k+1}, \dots, p_n) e^{-i \left[ \Sigma_j + p_k p_\w \right]} + 
    \\
    &\hphantom{=\sum_{j < k} \overset{\circ}{\Upsilon}{}_n^{[j]}(0|}
    + \overset{\circ}{\Upsilon}{}_n^{[k]}(0| p_\w + \pi_k; 0, \dots, 0, p_{k+1}, \dots, p_n) e^{-i \left[ \Sigma_k + \pi_k p_\w \right]}+
    \\
    &\hphantom{=\sum_{j < k} \overset{\circ}{\Upsilon}{}_n^{[j]}(0| \pi_j;0, \dots, 0,}
    +\sum_{j > k} \overset{\circ}{\Upsilon}{}_n^{[j]}(0|p_\w + \pi_j; 0, \dots, 0, p_{j+1}, \dots, p_n) e^{-i \left[ \Sigma_j + \pi_k p_\w + p_\w (\pi_j - \pi_k) \right]} -
    \\
    &-\sum_{j < k} \overset{\circ}{\Upsilon}{}_n^{[j]}(0| \pi_j; 0, \dots, 0, p_{j+1}, \dots, p_k, p_\w + p_{k+1}, \dots, p_n) e^{-i \left[ \Sigma_j + p_\w p_{k+1}  \right]} -
    \\
    &\hphantom{=\sum_{j < k} \overset{\circ}{\Upsilon}{}_n^{[j]}(0|}
    - \overset{\circ}{\Upsilon}{}_n^{[k]}(0| \pi_k; 0, \dots, 0, p_\w + p_{k+1}, \dots, p_n) e^{-i \left[ \Sigma_k + p_\w p_{k+1} \right]} - 
    \\
    &\hphantom{=\sum_{j < k} \overset{\circ}{\Upsilon}{}_n^{[j]}(0| \pi_j;0, \dots, 0,}
    - \sum_{j > k} \overset{\circ}{\Upsilon}{}_n^{[j]}(0|p_\w + \pi_j; 0, \dots, 0, p_{j+1}, \dots, p_n) e^{-i \left[ \Sigma_j + \pi_k p_\w + p_\w (\pi_j - \pi_k) \right]} =
\end{aligned} \\
\begin{aligned}
    = \overset{\circ}{\Upsilon}{}_n^{[k]}(0| p_\w + \pi_k; 0, \dots, 0, & p_{k+1}, \dots, p_n)  e^{-i \left[ \Sigma_k + \pi_k p_\w \right]} - \overset{\circ}{\Upsilon}{}_n^{[k]}(0| \pi_k; 0, \dots, 0, p_\w + p_{k+1}, \dots, p_n) e^{-i \left[ \Sigma_k + p_\w p_{k+1} \right]} + 
    \\
    + \sum_{j<k} e^{-i \Sigma_j} & \left( \overset{\circ}{\Upsilon}{}_n^{[j]}(0| \pi_j; 0, \dots, 0, p_{j+1}, \dots, p_k + p_\w, p_{k+1}, \dots, p_n) e^{-i  p_k p_\w }  \right.-
    \\
        & \quad -  \left. \overset{\circ}{\Upsilon}{}_n^{[j]}(0| \pi_j; 0, \dots, 0, p_{j+1}, \dots, p_k, p_\w + p_{k+1}, \dots, p_n) e^{-i  p_\w p_{k+1} } \right).
\end{aligned}
\end{multline}
Let us first study the sum over $j<k$ in the last equation, considering each summand separately. Setting $\pwo \rightarrow \pi_j, \; \pwt \rightarrow p_\w, \; p_1, \dots, p_j \rightarrow 0$ in \eqref{C^.wC^.wC^.} and taking into account the assumed requirement on one-from sector vertices \eqref{requirement_cyclic} (hence, the LHS of \eqref{C^.wC^.wC^.} vanishes) we obtain
\begin{multline}
\hphantom{\overset{\circ}{\Upsilon}{}_n^{[j]}(0|}
\begin{aligned}
    &\overset{\circ}{\Upsilon}{}_n^{[j]}(0| \pi_j; 0, \dots, 0, p_{j+1}, \dots, p_k + p_\w, p_{k+1}, \dots, p_n) e^{-i  p_k p_\w } -
    \\
    - & \overset{\circ}{\Upsilon}{}_n^{[j]}(0| \pi_j; 0, \dots, 0, p_{j+1}, \dots, p_k, p_\w + p_{k+1}, \dots, p_n) e^{-i  p_\w p_{k+1} } =
\end{aligned}
    \\
\begin{aligned}
    = & \overset{\circ}{\Upsilon} {}_n^{[k]}(0|p_\w; 0, \dots, 0, \pi_j, \dots, p_n) -
    \\
    - & \overset{\circ}{\Upsilon} {}_n^{[k]}(0|p_\w; 0, \dots, 0, 0, \pi_{j + 1}, \dots, p_n) e^{-i \pi_j p_{j + 1}}
\end{aligned}
\end{multline}
Consequently, the original sum over $j< k$ simplifies to
\begin{multline}
\begin{aligned}
    \sum_{j<k} e^{-i \Sigma_j} & \left( \overset{\circ}{\Upsilon}{}_n^{[j]}(0| \pi_j; 0, \dots, 0, p_{j+1}, \dots, p_k + p_\w, p_{k+1}, \dots, p_n) e^{-i  p_k p_\w }  \right.-
    \\
    & \quad -  \left. \overset{\circ}{\Upsilon}{}_n^{[j]}(0| \pi_j; 0, \dots, 0, p_{j+1}, \dots, p_k, p_\w + p_{k+1}, \dots, p_n) e^{-i  p_\w p_{k+1} } \right) =
\end{aligned}
    \\
    \sum_{j<k} e^{-i \Sigma_j} \left(  \overset{\circ}{\Upsilon} {}_n^{[k]}(0|p_\w; 0, \dots, 0, \pi_j, \dots, p_n)  -  \overset{\circ}{\Upsilon} {}_n^{[k]}(0|p_\w; 0, \dots, 0, 0, \pi_{j + 1}, \dots, p_n) e^{-i \pi_j p_{j + 1}} \right) =
    \\
    = \sum_{j<k}  \left(  \overset{\circ}{\Upsilon} {}_n^{[k]}(0|p_\w; 0, \dots, 0, \pi_j, \dots, p_n) e^{-i \Sigma_j} - \overset{\circ}{\Upsilon} {}_n^{[k]}(0|p_\w; 0, \dots, 0, 0, \pi_{j + 1}, \dots, p_n) e^{-i \Sigma_{j+1}} \right) = 
    \\
    = \overset{\circ}{\Upsilon} {}_n^{[k]}(0|p_\w; p_1, \dots, p_n)  - \overset{\circ}{\Upsilon} {}_n^{[k]}(0|p_\w; 0, \dots, 0, \pi_k, p_{k+1} \dots, p_n) e^{-i \Sigma_k}.
\end{multline}
Thereby, 
\begin{multline}
    \dd_x \delta B_n \big|_{y=0}^{C^k\w C^{n-k}} = \overset{\circ}{\Upsilon} {}_n^{[k]}(0|p_\w; p_1, \dots, p_n) +
    \\ 
    + \overset{\circ}{\Upsilon}{}_n^{[k]}(0| p_\w + \pi_k; 0, \dots, 0,  p_{k+1}, \dots, p_n)  e^{-i \left[ \Sigma_k + \pi_k p_\w \right]} - \overset{\circ}{\Upsilon}{}_n^{[k]}(0| \pi_k; 0, \dots, 0, p_\w + p_{k+1}, \dots, p_n) e^{-i \left[ \Sigma_k + p_\w p_{k+1} \right]}
    \\
    - \overset{\circ}{\Upsilon} {}_n^{[k]}(0|p_\w; 0, \dots, 0, \pi_k, p_{k+1} \dots, p_n) e^{-i \Sigma_k}
\end{multline}
It now remains to return to the compatibility condition of the unfolded nonlinear higher-spin equations in the form  \eqref{C^.wwC^.},  this time with the substitution $\pwo \rightarrow \pi_k, \; \pwt \rightarrow p_\w$, and $p_1, \dots, p_k \rightarrow 0$ (again, by virtue of the assumed cyclic symmetry of the one-form sector vertices, the LHS of \eqref{C^.wwC^.} is equal to zero):
\begin{multline}
     0 = \overset{\circ}{\Upsilon} {}_n^{[k]}(0|p_\w; 0, \dots, 0, \pi_k, p_{k+1} \dots, p_n) 
     -
     \overset{\circ}{\Upsilon}{}_n^{[k]}(0| p_\w + \pi_k; 0, \dots, 0,  p_{k+1}, \dots, p_n)  e^{-i \pi_k p_\w} 
     \\
     +
     \overset{\circ}{\Upsilon}{}_n^{[k]}(0| \pi_k; 0, \dots, 0, p_\w + p_{k+1}, \dots, p_n) e^{-i  p_\w p_{k+1} }.
\end{multline}
This proves that 
\begin{equation}
    \dd_x \delta B_n \big|_{y=0}^{C^k\w C^{n-k}} = \overset{\circ}{\Upsilon} {}_n^{[k]}(0|p_\w; p_1, \dots, p_n),
\end{equation}
so the shift \eqref{Bn_shift} resolves \eqref{Bn_equation} in all orderings up to the trailing one, namely $C^n \omega$, which we now consider.
\subsection{Trailing ordering: \texorpdfstring{$C^n \w$}{Cnw}}
To study this ordering it is convenient to assume that we have already performed shift \eqref{Bn_shift}, thus for all $k < n$
\begin{equation}
    \Upsilon {}_n^{[k]} (0| p_\w; p_1, \dots, p_n) = 0.
\end{equation}
It follows that, using \eqref{wwC^.} and \eqref{wC^.wC^.}, we can express $\Upsilon {}_n^{[k]}$ for all $k<n$ in terms of the one-form sector vertices as follows
\begin{equation}
    \Upsilon_n^{[k]} (y| p_\w; p_1, \dots, p_n) = \Upsilon_{n-1}^{[0, k]} ((-1)^n p_n|y,  p_\w ;p_1, \dots, p_{n-1}).
    \label{first_orderings_duality}
\end{equation}
Since the latter ones are fixed, it fixes  ${\Upsilon} {}_n^{[k]}$ for all $k<n$  uniquely. That is why the only leftover shift we can perform to $B_n$ must not change first $n-1$ orderings. Specifically, decomposing\footnote{The superscript "$(lo)$" in $\delta B_n^{(lo)}$ is not an index but an abbreviation for "leftover".} $\left(\dd_x \delta B_n^{(lo)} \big|^{\w C^n} + \left[\w, \delta B_n^{(lo)} \right]_* \right)\bigg|_{y=0}$ into different orderings we obtain the following conditions
\begin{align}
    \w C^n: \quad & \delta B_n^{(lo)}(p_\w| p_1, \dots, p_n) = \delta B_n^{(lo)}(0|p_\w + p_1, p_2, \dots, p_n) e^{-i p_\w p_1}; \\
    C^k \w C^{n-k}: \quad & \delta B_n^{(lo)}(0|p_1, \dots, p_k + p_\w, p_{k+1}, \dots, p_n) e^{-i p_k p_\w}  =  \delta B_n^{(lo)}(0|p_1, \dots, p_k , p_\w +  p_{k+1}, \dots, p_n) e^{-i p_\w p_{k+1} }.
\end{align}
It is straightforward to see (a short proof of this statement is given in Appendix \ref{leftover_shift}) that all fields $\delta B_n^{(lo)}$ satisfying these conditions for all $k<n$ have the form
\begin{equation}
    \delta B_n^{(lo)}(y|p_1, \dots, p_n) = \mathrm{const} \cdot e^{-i \left[y \pi_n + \Sigma_n \right]},
\end{equation}
which equals, up to a factor, $C*\dots*C k^{n+1}$. Here one  can see differences between odd- and even-$n$ cases. Indeed, when $n$ is odd, $k^{n+1} = 1$ and expression simplifies to $C*\dots * C$. Since $\dd_x C + \left[\omega, C \right]_* = 0$ up to higher terms in $C$, by virtue of the Leibniz  rule we have
\begin{equation}
    \dd_x C *\dots *C\big|^{\w C^n} + \left[\omega, C*\dots*C \right]_* = 0,
\end{equation} 
so the leftover shift does not change the vertex at all. It remains only to prove that, after performing the shift \eqref{Bn_shift}, the last ordering satisfies \eqref{U_(y=0)=0} too. 
On the other hand, for even $n$ shift $\delta B_n^{(lo)} = C*\dots*Ck$ nontrivially contributes to the last ordering
\begin{equation}
    \dd_x \delta B_n^{(lo)} \big|_{y=0}^{C^n \w} +  \delta B_n^{(lo)} * \w \big|_{y=0} = e^{-i \Sigma_n} \left(e^{i p_\w \pi_n} - e^{-i p_\w \pi_n} \right).
    \label{C...Ck_contribution}
\end{equation}
However, we can get some useful identities in both cases uniformly. Firstly, due to the equalities \eqref{first_orderings_duality} and \eqref{requirement_cyclic_2} the LHS of \eqref{C^.wC^.w} vanishes and so $\forall \;k: 0<k<n$
\begin{equation}
    \overset{\circ}{\Upsilon}{}_n^{[n]} (0|\pwt; p_1, \dots, p_k + \pwo, p_{k+1}, \dots, p_n) e^{-i p_k \pwo} = \overset{\circ}{\Upsilon}{}_n^{[n]} (0|\pwt; p_1, \dots, p_k , \pwo + p_{k+1}, \dots, p_n) e^{-i  \pwo p_k }.
\end{equation}
One can set $\pwt \rightarrow p_\w, \; p_1 \rightarrow 0, \pwo \rightarrow p_1$ in $k=1$ case and get 
\begin{equation}
    \overset{\circ}{\Upsilon}{}_n^{[n]} (0|p_\w; p_1, \dots, p_n) = \overset{\circ}{\Upsilon}{}_n^{[n]} (0|p_\w;0, p_1 + p_2, \dots, p_n) e^{-i p_1 p_2 }.
\end{equation}
Analogously, making various variable substitution one derives
\begin{equation}
    \overset{\circ}{\Upsilon}{}_n^{[n]} (0|p_\w; p_1, \dots, p_n) = \overset{\circ}{\Upsilon}{}_n^{[n]} (0|p_\w;0, p_1 + p_2, \dots, p_n) e^{-i p_1 p_2 } = 
    \dots =
     \overset{\circ}{\Upsilon}{}_n^{[n]} (0|p_\w; 0, \dots, 0, \pi_n) e^{-i \Sigma_n}
     \label{U_n^n_reduction}
\end{equation}
and 
\begin{equation}
    \overset{\circ}{\Upsilon}{}_n^{[n]} (0|p_\w; p, 0, \dots, 0) = \overset{\circ}{\Upsilon}{}_n^{[n]} (0|p_\w; 0, p, \dots, 0) = \dots = \overset{\circ}{\Upsilon}{}_n^{[n]} (0|p_\w; 0, \dots, 0, p).
    \label{U_n^n_cyclic}
\end{equation}
Hereby, the intended property \eqref{U_(y=0)=0} in the last ordering is equivalent to much simpler ones 
\begin{equation}
    \overset{\circ}{\Upsilon}{}_n^{[n]} (0|p_\w; p_1, \dots, p_n) = 0 \Leftrightarrow \overset{\circ}{\Upsilon}{}_n^{[n]} (0|p_\w; p, 0, \dots, 0) = 0 \Leftrightarrow \overset{\circ}{\Upsilon}{}_n^{[n]} (0|p_\w; 0, \dots, 0, p) = 0.
\end{equation}
In order to move forward, let us consider equations \eqref{C^.ww} and \eqref{wC^.w} with the substitutions $\pwo \rightarrow p_\w, \; \pwt \rightarrow p, \; p_2, \dots, p_n \rightarrow 0$:
\begin{multline}
    -\Upsilon_{n-1}^{[n-1, n-1]}(p_1|p_\w, p; 0, \dots, 0) - \overset{\circ}{\Upsilon}{}_n^{[n]}((-1)^n \, p| p_\w; p_1, 0, \dots, 0) = 
    \\
    = \overset{\circ}{\Upsilon}{}_n^{[n]} (0| p; p_1, 0, \dots, 0, p_\w) - \overset{\circ}{\Upsilon}{}_n^{[n]}(0|p_\w + p; p_1, 0, \dots, 0) e^{-i p_\w p};
    \label{C^nw_base_1}
\end{multline}
\begin{equation}
    -\overset{\circ}{\Upsilon}{}_n^{[n]}(p_\w| p; p_1, 0, \dots, 0) - \Upsilon_{n-1}^{[n-1, n-1]}((-1)^n \, p_1| (-1)^n \, p, p_\w; 0, \dots, 0) 
    = - \overset{\circ}{\Upsilon}{}_n^{[n]}(0| p; p_\w + p_1, 0, \dots, 0) e^{-i p_\w p_1} .
    \label{C^nw_base_2}
\end{equation}
Here we used that, due to \eqref{first_orderings_duality}, \eqref{requirement_cyclic},
\begin{multline}
    \Upsilon_n^{[0]}((-1)^n \, p| p_\w; p_1, \dots, 0) = \Upsilon_{n-1}^{[0, 0]}(0| (-1)^n \, p, p_\w; p_1, 0, \dots, 0) = 
    \\
    = \Upsilon_{n-1}^{[1, 1]}(0| (-1)^n \, p, p_\w; 0, p_1, 0, \dots, 0) = \dots = \Upsilon_{n-1}^{[n-1, n-1]}((-1)^n \, p_1| (-1)^n \, p, p_\w; 0, \dots, 0) .
\end{multline}

\subsubsection{Odd-n sector}
 In this sector equations \eqref{C^nw_base_1}, \eqref{C^nw_base_2} simplify and we can set $p_1 \rightarrow 0$: 
\begin{multline}
    -\Upsilon_{n-1}^{[n-1, n-1]}(0|p_\w, p; 0, \dots, 0) - \overset{\circ}{\Upsilon}{}_n^{[n]}(- p| p_\w; 0, \dots, 0) = 
    \\
    = \overset{\circ}{\Upsilon}{}_n^{[n]} (0| p; 0, \dots, 0, p_\w) - \overset{\circ}{\Upsilon}{}_n^{[n]}(0|p_\w + p; 0, \dots, 0) e^{-i p_\w p};
    \label{C^nw_base_odd_1}
\end{multline}
\begin{equation}
    -\overset{\circ}{\Upsilon}{}_n^{[n]}(p_\w| p; 0, \dots, 0) - \overset{\circ}{\Upsilon}{}_{n-1}^{[n-1, n-1]}(0| - p, p_\w; 0, \dots, 0) 
    = - \overset{\circ}{\Upsilon}{}_n^{[n]}(0| p; p_\w, 0, \dots, 0) .
    \label{C^nw_base_odd_2}
\end{equation}
A crucial observation is that all functions of the spinor variables considered herein are $\mathrm{sp}(2)$-invariant. This implies that if a function depends on the spinors $p_1, p_2, \dots, p_n$, they enter the expression 
exclusively through the $\mathrm{sp}(2)$-contractions $p_i p_j$. However, since $p_i p_j = - p_j p_i$, then any function $F(p_1, p_2)$  obeys
\begin{equation}
    F(p_1, p_2) = F(p_2, -p_1) = F(-p_2, p_1) = F(-p_1, -p_2).
\end{equation}
Specifically,
\begin{equation}
\begin{aligned}
    \Upsilon_{n-1}^{[n-1, n-1]}(0| - p, p_\w; 0, \dots, 0) &= \Upsilon_{n-1}^{[n-1, n-1]}(0|p_\w, p; 0, \dots, 0), 
    \\
    \overset{\circ}{\Upsilon}{}_n^{[n]}(p_\w| p; 0, \dots, 0) &= \overset{\circ}{\Upsilon}{}_n^{[n]}(- p| p_\w; 0, \dots, 0).
\end{aligned}
\end{equation}
That is why LHSs of equations \eqref{C^nw_base_odd_1} and \eqref{C^nw_base_odd_2} coincide which brings us to
\begin{equation}
    - \overset{\circ}{\Upsilon}{}_n^{[n]}(0| p; p_\w, 0, \dots, 0) =  \overset{\circ}{\Upsilon}{}_n^{[n]} (0| p; 0, \dots, 0, p_\w) - \overset{\circ}{\Upsilon}{}_n^{[n]}(0|p_\w + p; 0, \dots, 0)e^{-i p_\w p}.
\end{equation}
The fact that $(p_\w + p)(p_\w + p) = 0$ and \eqref{U_n^n_cyclic} yields
\begin{equation}
    \overset{\circ}{\Upsilon}{}_n^{[n]} (0| p_\w; p, 0, \dots, 0) = \frac{1}{2} \overset{\circ}{\Upsilon}{}_n^{[n]}(0|p_\w + p; 0, \dots, 0) e^{-i p_\w p} = -\frac{1}{2} \overset{\circ}{\Upsilon}{}_n^{[n]}(0|0; 0, \dots, 0) e^{-i p_\w p} = 0.
\end{equation}
Here last equality holds true as a consequence of \eqref{C^nw_base_odd_1} with $p, p_\w \rightarrow 0$ and imposed restriction \eqref{requirement_nulling}. This completes the construction of the holomorphic vertex $\Upsilon(\w C^n)$ satisfying property \eqref{U_(y=0)=0} in the odd-$n$ case. It is worth noting here that in this instance the field $B_n$ leading to the vertex $\Upsilon(\w C^n)$ with the desired property \eqref{U_(y=0)=0} is not unique, but is determined up to a shift by term proportional to $C*\dots*C$. A detailed analysis of the explicit expression for the resulting vertex will be deferred until after we have considered the even-$n$ case, to which we now turn.
\subsubsection{Even-n sector}
 In this sector \eqref{C^nw_base_1}, \eqref{C^nw_base_2} take the form (in view of \eqref{U_n^n_cyclic})
\begin{multline}
    -\Upsilon_{n-1}^{[n-1, n-1]}(p_1|p_\w, p; 0, \dots, 0) - \overset{\circ}{\Upsilon}{}_n^{[n]}(p| p_\w; p_1, 0, \dots, 0) = 
    \\
    = \overset{\circ}{\Upsilon}{}_n^{[n]} (0| p; p_1 + p_\w, 0, \dots, 0) e^{-i p_1 p_\w} - \overset{\circ}{\Upsilon}{}_n^{[n]}(0|p_\w + p; p_1, 0, \dots, 0) e^{-i p_\w p};
    \label{C^nw_base_even_1}
\end{multline}
\begin{equation}
    -\overset{\circ}{\Upsilon}{}_n^{[n]}(p_\w| p; p_1, 0, \dots, 0) - \Upsilon_{n-1}^{[n-1, n-1]}(p_1| p, p_\w; 0, \dots, 0) 
    = - \overset{\circ}{\Upsilon}{}_n^{[n]}(0| p; p_\w + p_1, 0, \dots, 0) e^{-i p_\w p_1},
    \label{C^nw_base_even_2}
\end{equation}
 so setting $p_1$ to zero does not allow us to get meaningful results. This is why we must first differentiate\footnote{The author is grateful to Vyacheslav Didenko for suggesting the idea of using derivatives in this context.} with respect to $p_1$, and only afterwards substitute $p_1 \rightarrow 0$. However, we initially need to interchange $p_\omega \leftrightarrow p$ in \eqref{C^nw_base_even_2}. After this replacement, the LHSs of \eqref{C^nw_base_even_1} and \eqref{C^nw_base_even_2} coincide, so
 \begin{equation}
     - \overset{\circ}{\Upsilon}{}_n^{[n]}(0| p_\w; p + p_1, 0, \dots, 0) e^{-i p p_1} = \overset{\circ}{\Upsilon}{}_n^{[n]} (0| p; p_1 + p_\w, 0, \dots, 0) e^{-i p_1 p_\w} - \overset{\circ}{\Upsilon}{}_n^{[n]}(0|p_\w + p; p_1, 0, \dots, 0) e^{-i p_\w p}.
 \end{equation}
Therefore, applying $q^\alpha \frac{\partial}{\partial p_1^\alpha}$, where $q$ is an auxiliary spinor variable, and setting $p_1 \rightarrow 0$ brings us to
\begin{multline}
    i \,pq \overset{\circ}{\Upsilon}{}_n^{[n]}(0| p_\w; p , 0, \dots, 0)  - q^\a \left(\partial_\a \overset{\circ}{\Upsilon}{}_n^{[n]} \right)(0| p_\w; p, 0, \dots, 0) =
    \\
    = i \, p_\w q \overset{\circ}{\Upsilon}{}_n^{[n]} (0| p; p_\w, 0, \dots, 0) +  q^\a \left(\partial_\a \overset{\circ}{\Upsilon}{}_n^{[n]} \right) (0| p; p_\w, 0, \dots, 0) - q^\a \left(\partial_\a \overset{\circ}{\Upsilon}{}_n^{[n]} \right)(0|p_\w + p; 0, \dots, 0) e^{-i p_\w p}.
\end{multline}
Here $\partial_\a \overset{\circ}{\Upsilon}{}_n^{[n]}$ denotes partial derivative of $\overset{\circ}{\Upsilon}{}_n^{[n]}$ with respect to $p_1$ argument. Subsequently, decomposing $e^{-i p_\w p}$ into sum of sine and cosine function we obtain
\begin{multline}
    i \left( pq \overset{\circ}{\Upsilon}{}_n^{[n]}(0| p_\w; p , 0, \dots, 0) - p_\w q \overset{\circ}{\Upsilon}{}_n^{[n]} (0| p; p_\w, 0, \dots, 0) - q^\a \left(\partial_\a \overset{\circ}{\Upsilon}{}_n^{[n]} \right)(0|p_\w + p; 0, \dots, 0) \sin (p p_\w)\right) =
    \\
    q^\a \left(\partial_\a \overset{\circ}{\Upsilon}{}_n^{[n]} \right) (0| p; p_\w, 0, \dots, 0) + q^\a \left(\partial_\a \overset{\circ}{\Upsilon}{}_n^{[n]} \right)(0| p_\w; p, 0, \dots, 0) - q^\a \left(\partial_\a \overset{\circ}{\Upsilon}{}_n^{[n]} \right)(0|p_\w + p; 0, \dots, 0) \cos (p_\w p).
    \label{odd-even_C^nw_eq}
\end{multline}
One can notice that the RHS of \eqref{odd-even_C^nw_eq} is symmetric under $p \leftrightarrow p_\omega$, whereas the LHS is antisymmetric under the same interchange. It entails that both LHS and RHS of \eqref{odd-even_C^nw_eq} vanish independently. Considering, for example, LHS with $q \rightarrow p_\w$ one gets
\begin{equation}
    pp_\w \overset{\circ}{\Upsilon}{}_n^{[n]}(0| p_\w; p , 0, \dots, 0) = p_\w^\a \left(\partial_\a \overset{\circ}{\Upsilon}{}_n^{[n]} \right)(0|p_\w + p; 0, \dots, 0) \sin (p p_\w).
\end{equation}
But since all contractions $(p_\w + p)(p_\w + p)$ vanish, the expression $\left(\partial_\a \overset{\circ}{\Upsilon}{}_n^{[n]} \right)(0|p_\w + p; 0, \dots, 0)$ necessarily has the form
\begin{equation}
    \left(\partial_\a \overset{\circ}{\Upsilon}{}_n^{[n]} \right)(0|p_\w + p; 0, \dots, 0) = A (p_\w + p)_\a,
\end{equation}
where $A$ is some constant. Thereby,
\begin{equation}
    pp_\w \overset{\circ}{\Upsilon}{}_n^{[n]}(0| p_\w; p , 0, \dots, 0) = - 2iA \; p p_\w \left(e^{i p_\w p} - e^{-i p_\w p} \right),
\end{equation}
which, after combining with \eqref{U_n^n_reduction} and \eqref{U_n^n_cyclic}, directly implies
\begin{equation}
    \overset{\circ}{\Upsilon}{}_n^{[n]}(0|p_\w; p_1, \dots, p_n) = \overset{\circ}{\Upsilon}{}_n^{[n]}(0|p_\w; \pi_n, 0, \dots, 0) e^{-i \Sigma_n} = -2i Ae^{-i \Sigma_n}\left(e^{i p_\w \pi_n} - e^{-i p_\w \pi_n} \right)
\end{equation}
Comparing this with \eqref{C...Ck_contribution} it is straightforward to see, that shift 
\begin{equation}
    \delta B_n^{(lo)} = -2iA \;C*\dots *C
\end{equation}
brings the vertex $\Upsilon(\w C^n)$ to a form that meets the condition \eqref{U_(y=0)=0} we wanted. 

\subsection{Interim result}
 Let us briefly repeat what we have managed to do and what are the consequences of this. We have considered $n$-th order of perturbation theory and assumed that holomorphic vertices $\Upsilon(\w C^k)$ for all $k<n$ are found in a spinor spin-local form with the additional feature
 \begin{equation}
     \Upsilon(\w C^k) \big|_{y=0} = 0 \qquad \forall k<n.
 \end{equation}
We also assumed that conditions \eqref{requirement_cyclic}--\eqref{requirement_nulling} are satisfied for the holomorphic vertex $\Upsilon(\w^2 C^{n-1})$ of the one-form sector. It allows us to find $B_n$, leading to holomorphic vertex $\Upsilon(\w C^n)$, such that $\Upsilon(\w C^n) \big|_{y=0} = 0$. Hence, a straightforward application of \eqref{wwC^.}, \eqref{C^.wwC^.} and \eqref{C^.ww} yields
\begin{align}
    \Upsilon_n^{[k]} (y| p_\w; p_1, \dots, p_n) &= \hphantom{-} \Upsilon_{n-1}^{[0, k]} ((-1)^n p_n|y,  p_\w ;p_1, \dots, p_{n-1}), \quad k<n;
    \label{local_form_k<n}
    \\
    \Upsilon_n^{[n]}(y| p_\w; p_1, \dots, p_n) &= -\Upsilon_{n-1}^{[n-1, n-1]}(p_1|p_\w, (-1)^n \, y; p_2, \dots, p_{n-1}).
    \label{local_form_n}
\end{align}
Thereby, if $\Upsilon_n^{[k_1, k_2]}(y|\pwo, \pwt; p_1, \dots, p_{n-1})$ is ultra-local, i.e. it does not contain contractions $y p_i$ and $p_i p_j$ ($i = 1, \dots, n-1$) in its exponents, than found $\Upsilon(\w C^n)$ is spinor spin-local. Together with \eqref{U_(y=0)=0} spinor spin-locality give rise \cite{proj_comp} to space-time spin-locality of $\Upsilon(\w C^n)$ vertex. 

An interesting aspect is that the situation is different for odd and even $n$. Indeed, for even-$n$ case the field $B_n$ leading to vertex with target property \eqref{U_(y=0)=0} is uniquely defined. However, there is a freedom in shift up to terms proportional to $C*\dots*C$ in odd-$n$ case. This shift is clearly nonlocal, but it can cancel the nonlocal part of $\overset{\circ}{B_n}$. Furthermore, the choice of $B_n$ can influence the feasibility of constructing an ultra-local vertex $\Upsilon(\w^2 C^n)$ in the one-form sector possessing cyclic symmetry \eqref{requirement_cyclic}, \eqref{requirement_cyclic_2} at the next order. We leave this issue open for the future exploration. 
Moreover, this $C*\dots*C$ shift can be absorbed by a redefinition of the master field $B \to B + a \cdot B*\dots*B$, which deforms the original equation \eqref{master_eqs_5} in the way proposed in \cite{nonlinear_system} and, at the considered $n$-th order, coincides with the shift $C \to C + a \cdot C*\dots*C$. Here $a$ is an arbitrary coefficient. 
For even $n$, the same redefinition contributes at $n$-th order only to the topological sector of the system and does not affect the interaction of the dynamical fields.

\subsection{Lower-order vertices}
As mentioned previously, the holomorphic vertices $\Upsilon(\omega^2 C)$ and $\Upsilon(\omega^2 C^2)$, computed in \cite{w2C} and \cite{1909} respectively, do indeed obey the assumptions \eqref{requirement_cyclic}-\eqref{requirement_nulling}, which can be checked by a straightforward computation. Moreover, these vertices are ultra-local. This implies that the proposed scheme leads to spin-local holomorphic vertices $\Upsilon(\omega C^2)$ and $\Upsilon(\omega C^3)$. As for the $\Upsilon(\omega C^2)$ vertex, previously found in a spin-local form in \cite{wCC_local},  this was achieved by choosing
\begin{equation}
    \overset{\circ}{B_2} = \frac{\eta}{4i} C*C*\Delta_{p_1 + 2p_2} \Delta_{p_2} \gamma,
    \label{B_2_sh}
\end{equation}
where \cite{Homotopy_Operators}
\begin{equation}
    \Delta_q F(z, y|\theta) = \int_0^1 \frac{d\tau}{\tau} (z + q)^\a \frac{\partial}{\partial \theta^\a} F(\tau z - (1-\tau)q, y| \tau \theta), 
\end{equation}
and then manually shifting it
\begin{align}
    B_2 & = \overset{\circ}{B_2} + \delta B_2, 
    \\
    \delta B_2 &= \frac{\eta}{2} \int_0^1 d\tau C(\tau y, \bar{y}, K) \bar{*} C((\tau -1)y, \bar{y} , K)k.
    \label{delta_B_2}
\end{align}
This shift leads to a spinor spin-local holomorphic vertex that vanishes at $y=0$. Thus, as stated above in the discussion of the even-$n$ case, there is only one $B_2$ leading to such a vertex. That is why the algorithm described in this paper, applied to $\overset{\circ}{B}_2$ \eqref{B_2_sh}, reproduces the same shift $\delta B_2$ \eqref{delta_B_2}.

Nevertheless, the spin-local form of the vertex $\Upsilon(\omega C^3)$ is of much greater interest, as it was not known previously. In \cite{quadratic_spin-locality}, it was proven that $\Upsilon(\omega C^3)$ admits a spinor spin-local form, although it was explicitly given only in one ordering in \cite{manifest_form}. However, nothing was known about the projective compactness \cite{proj_comp} of this vertex, without which it could fail to be space-time spin-local despite being spinor spin-local. In fact, this work proves that the holomorphic vertex $\Upsilon(\omega C^3)$ is not only spinor spin-local, but also projectively compact. The concrete expression for this vertex can be found by virtue of \eqref{local_form_k<n} and \eqref{local_form_n} (see Appendix \ref{explicit_form}).

\section{Construction of arbitrary vertices in the one-form sector}
\label{construction}
Here we set another task, namely, we again consider, by induction, the $n$-th order of perturbations theory in powers of the $C$-field. This means that the auxiliary fields $S_1, \dots, S_{n-1}$, $W_0 = \w(y), W_1, \dots, W_{n-1}$, and $B_1 = C(y), B_2, \dots, B_{n-1}$ are found and fixed. And let us also assume, that we are given consistent holomorphic vertex $\Upsilon(\w C^3)$. We need to find $B_n$, such that
\begin{equation}
     \Upsilon(\w C^n) = - \sum_{j=2}^{n} \dd_x B_j \big|^{\w C^n} - \sum_{j=0}^{n-1}\left[W_j, B_{n-j} \right]_*.
 \end{equation}
Acting analogously we find any particular solution $\overset{\circ}{B_n}$, leading to $\overset{\circ}{\Upsilon}(\w C^n)$. Then, our goal is to obtain $\delta B_n(y)$, satisfying 
\begin{equation}
         \dd_x \delta B_n \big|^{\w C^n} + \left[\w, \delta B_n \right]_* =  \overset{\circ}{\Upsilon}(\w C^n) - \Upsilon(\w C^n) =: F,
         \label{F_definition}
\end{equation}
where we denoted the RHS of the last equation as $F$. Let us write down this expression in concrete orderings:
\begin{equation}
    \delta B_n(y+ p_\w| p_1, \dots, p_n) e^{-i y p_\w} - \delta B_n(y|p_\w + p_1, p_2, \dots, p_n) e^{-i p_\w p_1} = F^{[0]}(y|p_\w; p_1, \dots, p_n);
    \label{arbitrary_vertex_eq_1}
\end{equation}
\begin{multline}
    \delta B_n(y| p_1, \dots, p_k + p_\w, p_{k+1}, \dots, p_n) e^{-i p_k p_\w} - \delta B_n(y| p_1 , \dots, p_k, p_\w+ p_{k+1}, \dots, p_n) e^{-i p_k p_\w} e^{-i p_\w p_{k+1}} =
    \\
    = F^{[k]}(y|p_\w; p_1, \dots, p_n), \qquad 0<k<n;
    \label{arbitrary_vertex_eq_k}
\end{multline}
\begin{equation}
    \delta B_n(y| p_1, \dots, p_{n-1}, p_n + p_\w ) e^{-i p_n p_\w} - \delta B_n(y+ (-1)^n p_\w| p_1, \dots, p_n) e^{-i (-1)^n p_\w y} = F^{[n]}(y|p_\w; p_1, \dots, p_n);
    \label{arbitrary_vertex_eq_n}
\end{equation}
where $F^{[k]}(y|p_\w; p_1, \dots, p_n)$ stays for $C^k\w C^{n-k}$ ordering of $F$.

Exploiting  \eqref{arbitrary_vertex_eq_1} with substitutions $y\rightarrow 0, \, p_\w \rightarrow y$ and \eqref{arbitrary_vertex_eq_k} with substitutions (recall the notation for $\pi_k$ and $\Sigma_k$ in \eqref{p_notations}) $y \rightarrow 0, \, p_\w \rightarrow y + \pi_k, \, p_1, \dots, p_k \rightarrow 0$ we obtain
\begin{multline}
    \delta B_n(y|p_\w; p_1, \dots, p_n) =
    \\
    = \sum_{k=0}^{n-1}  F^{[k]}(0|y + \pi_k; 0, \dots, 0, p_{k+1}, \dots, p_n) e^{-i \left[y\pi_k + \Sigma_k \right]} + \delta B_n(0|0, \dots,0,y + \pi_n) e^{-i \left[y \pi_n + \Sigma_n \right]} =
    \\
    = \sum_{k=0}^{n-1}  F^{[k]}(0|y + \pi_k; 0, \dots, 0, p_{k+1}, \dots, p_n) e^{-i \left[y\pi_k + \Sigma_k \right]} +  A e^{-i \left[y \pi_n + \Sigma_n \right]}.
    \label{arbitrary_vertex_gen_solution}
\end{multline}
Here, $\delta B_n$ must be $\mathrm{sp}(2)$-invariant, so $B_n(0|0, \dots,0,y + \pi_n) = B_n(0|0, \dots,0, 0) = A$ for some constant $A$. It is then straightforward to see that \eqref{arbitrary_vertex_gen_solution} provides a general solution to equations \eqref{arbitrary_vertex_eq_1} and \eqref{arbitrary_vertex_eq_k}.  It remains to require that it also satisfies \eqref{arbitrary_vertex_eq_n}. In odd-$n$ case it forces $F$ to fulfill
\begin{multline}
    F^{[n]}(y|p_\w; p_1, \dots, p_n) = \sum_{k=0}^{n-1} \left( F^{[k]}(0|y + \pi_k; 0, \dots, 0, p_{k+1}, \dots, p_n + p_\w) e^{-i \left[y\pi_k + \Sigma_k + p_n p_\w \right]} - \right.
    \\
    - \left.F^{[k]}(0|y - p_\w + \pi_k; 0, \dots, 0, p_{k+1}, \dots, p_n) e^{-i \left[y\pi_k + \Sigma_k + p_\w \pi_k + y p_\w  \right]} \right).
    \label{arbitrary_vertex_odd_requirement}
\end{multline}
In even-$n$ case the same equations \eqref{arbitrary_vertex_eq_n} requires
\begin{multline}
    F^{[n]}(y|p_\w; p_1, \dots, p_n) - \sum_{k=0}^{n-1} \left( F^{[k]}(0|y + \pi_k; 0, \dots, 0, p_{k+1}, \dots, p_n + p_\w) e^{-i \left[y\pi_k + \Sigma_k + p_n p_\w \right]} - \right.
    \\
    - \left.F^{[k]}(0|y + p_\w + \pi_k; 0, \dots, 0, p_{k+1}, \dots, p_n) e^{-i \left[y\pi_k + \Sigma_k + p_\w \pi_k +p_\w y \right]} \right)
    \label{arbitrary_vertex_even_requirement}
\end{multline}
to be proportional to $e^{-i \left[y\pi_n + \Sigma_n \right]} \sin \left((y + \pi_n) p_\w \right)$. Since $F$ can be expressed in terms of $\Upsilon(\omega C^n)$ through
the relation \eqref{F_definition}, equations
\eqref{arbitrary_vertex_odd_requirement} and
\eqref{arbitrary_vertex_even_requirement} translate into conditions on
$\Upsilon(\omega C^n)$. These conditions must be met for the vertex
$\Upsilon(\w C^n)$ to be realizable with some $B_n$, while $S_k$, $W_k$
and $B_k$ for $k<n$ are held fixed. Provided these conditions on
$\Upsilon(\w C^n)$ are satisfied, \eqref{arbitrary_vertex_gen_solution}
yields the required $B_n$ for arbitrary $A$ when $n$ is odd, and for a
specified value of $A$ when $n$ is even, the latter obtained by a direct
evaluation of \eqref{arbitrary_vertex_even_requirement}.
\section{Conclusion}
\label{sec:Conclusion}
In this work, our primary objective was to extract, from the original generating system \cite{nonlinear_system}, the vertices of the holomorphic zero-form sector of HS theory that satisfy the requirement \eqref{U_(y=0)=0}. The motivation for imposing this condition is as follows. Should the vertices additionally be shown to be spinor spin-local, then, together with \eqref{U_(y=0)=0}, this would imply their projective compactness and, consequently, their space-time spin-locality. However, \eqref{U_(y=0)=0} is not the only means of obtaining projectively compact vertices \cite{proj_comp}. Rather, it represents a more general condition. In practice, this broader condition proves easier to track and satisfy. Moreover, all holomorphic zero-form vertices derived from the holomorphic generating system \cite{Didenko} are found to comply with \eqref{U_(y=0)=0}.

We have shown that the necessary and sufficient conditions for existence of zero-from vertices fulfilling \eqref{U_(y=0)=0} are imposed on the holomorphic one-form sector vertices and are of two types, namely cyclic symmetry conditions \eqref{requirement_cyclic}, \eqref{requirement_cyclic_2} and a specific prefactor condition \eqref{requirement_nulling}. The necessity of these conditions is demonstrated in Appendix \ref{necessity}. We observe that the lower-order one-form sector vertices obtained from the original full nonlinear generating system \cite{w2C, 1909}, as well as the all-order one-form sector vertices derived from the holomorphic nonlinear generating system \cite{Povarnin}, satisfy both the cyclic symmetry and prefactor requirements. We therefore anticipate that these properties persist to all orders within the original system as well. Nevertheless, a promising direction for future research is to develop an explicit construction of one-form sector vertices exhibiting this structure. 

More concretely, our inductive scheme proceeds order by order in perturbation theory. At the $n$-th order, we assume that all holomorphic vertices $\Upsilon(\w C^k)$ with $k<n$ satisfy \eqref{U_(y=0)=0}. We further assume that the one-form sector vertex $\Upsilon(\w^2 C^{n-1})$ satisfies the cyclic symmetry and prefactor conditions \eqref{requirement_cyclic}--\eqref{requirement_nulling}. Under these assumptions, the scheme determines $B_n$ and yields a holomorphic vertex $\Upsilon(\w C^n)$ that again satisfies \eqref{U_(y=0)=0}. The resulting vertices at order $n$ are then expressed directly in terms of the $\Upsilon(\w^2C^{n-1})$ via \eqref{local_form_k<n} and \eqref{local_form_n}.
Furthermore, if the one-form sector vertices are ultralocal, then the zero-form vertices computed by the scheme described in this paper are space-time spin-local, which provides a promising avenue for obtaining local holomorphic vertices to all orders directly from the full generating system \cite{nonlinear_system}.

As a concrete application, we have implemented this mechanism to evaluate the spin-local holomorphic $\eta^2$ zero-form sector vertex.
This extends the results of \cite{quadratic_spin-locality, manifest_form}, where only the spinor spin-locality of this vertex had been established, by promoting it to space-time spin-locality.

Finally, an important structural point should be added concerning the uniqueness of the resulting vertices. For fixed holomorphic one-form sector vertices, the zero-form vertices satisfying requirement \eqref{U_(y=0)=0} are defined uniquely (that is in agreement with the vertex duality \cite{Povarnin}). However, while for odd $n$ the master field $B_n$ is also determined uniquely, for even $n$ there arises a one-parameter freedom at each step of the inductive procedure in perturbation theory, which must be fixed by an additional prescription. The choice of this parameter does not alter the vertex $\Upsilon(\w C^n)$ itself, but it can affect higher-order vertices in both the zero- and one-form sectors. Hence, it remains an open question how exactly to choose $B_n$ in the even-$n$ case, and we leave this subtlety for future investigation.

\section*{Acknowledgments} 
The author wishes to thank Misha Povarnin for inspiring the present investigation by recommending the study of vertex dualities, a notion first introduced in \cite{Povarnin}, and for his careful reading of the manuscript. Thanks also go to Vyacheslav Didenko for his valuable suggestion that helped overcome a particular technical obstacle, and to Mikhail Vasiliev and Olga Gelfond for many useful discussions. Vasiliev is additionally thanked for his careful reading and comments on this text.

\begin{appendices}

    \addtocontents{toc}{\protect\setcounter{tocdepth}{1}}
    \section{Useful formulas}
    \label{useful_formulas}
Here we provide star-multiplications formulas with $z$-independent functions, that are heavily used along this paper:
\begin{align}
    F(z; y) k^n * \varphi(y) &= \int d^2 p_\varphi \; d^2 r_\varphi e^{-i p_\varphi r_\varphi} F(z - (-1)^n p_\varphi; y - (-1)^n p_\varphi) \bar{*} \varphi(r_\varphi) k^n, \\
    \varphi(y) * F(z; y) k^n  &= \int d^2 p_\varphi \; d^2 r_\varphi e^{-i p_\varphi r_\varphi} \varphi(r_\varphi) \bar{*} F(z - p_\varphi; y + p_\varphi) k^n,
\end{align}
with implicit dependence on $\bar{y}$ and $\bar{k}$.
    \section{Necessity of the requirements imposed on the one-form sector vertices}
    \label{necessity}
Let us assume, that there exists holomorphic vertex $\Upsilon(\w C^n)$, such that $\Upsilon(\w C^n) \big|_{y=0} = 0$. Then, we infer from \eqref{C^.wwC^.} and \eqref{C^.wC^.wC^.} that for all $0<j<k<n$
\begin{align}
    \Upsilon_{n-1}^{[k, k]}((-1)^n p_n| \pwo, \pwt; p_1, \dots, p_{n-1}) &= \Upsilon_{n-1}^{[k-1, k-1]}(p_1|\pwo, \pwt; p_2, \dots, p_n), \\
    \Upsilon_{n-1}^{[j,k]}((-1)^n p_n| \pwo, \pwt; p_1, \dots, p_{n-1}) &= \Upsilon_{n-1}^{[j - 1,k - 1]}(p_1| \pwo, \pwt; p_2, \dots, p_n).
\end{align}
This proves \eqref{requirement_cyclic} for all $0< k_1 \leq k_2 < n$. Furthermore, combing \eqref{wC^.wC^.} and \eqref{C^.wC^.w}, we arrive at
\begin{multline}
    \Upsilon_{n-1}^{[0, k]} ((-1)^n p_n|(-1)^n \pwt,  \pwo;p_1, \dots, p_{n-1}) = \Upsilon_n^{[k]} ((-1)^n \pwt| \pwo; p_1, \dots, p_n) =
    \\
    = -\Upsilon_{n-1}^{[k-1, n-1]}(p_1|\pwo, \pwt; p_2, \dots , p_n) .
    \label{Appendix_cyclic_jump}
\end{multline}
One can notice that \eqref{Appendix_cyclic_jump} coincides with \eqref{requirement_cyclic_2}. 
Hence, we have shown that the requirements \eqref{requirement_cyclic} and \eqref{requirement_cyclic_2} are obligatory conditions for the existence of a consistent holomorphic vertex $\Upsilon(\omega C^n)$ satisfying $\Upsilon(\omega C^n)\big|_{y=0} = 0$. 

Besides, equations \eqref{wwC^.} and \eqref{wC^.wC^.} give
\begin{align}
    \Upsilon_{n-1}^{[0, 0]} ((-1)^n p_n|\pwo,  \pwt;p_1, \dots, p_{n-1}) &= \Upsilon_n^{[0]} (\pwo| \pwt; p_1, \dots, p_n) ,
    \\
    \Upsilon_{n-1}^{[0, k]} ((-1)^n p_n|\pwo,  \pwt;p_1, \dots, p_{n-1}) &= \Upsilon_n^{[k]} (\pwo| \pwt; p_1, \dots, p_n).
\end{align}
Since $\Upsilon_n^{[k]} (0| 0; p_1, \dots, p_n) = 0$ for all $k<n$ by our assumption, we have 
\begin{equation}
    \Upsilon_{n-1}^{[0, k]} ((-1)^n p_n|0, 0; p_1, \dots, p_{n-1}) = 0, \qquad \forall k<n.
\end{equation}
By the already established cyclic symmetry properties \eqref{requirement_cyclic} and \eqref{requirement_cyclic_2}, all other orderings of $\Upsilon(\omega^2 C^{n-1})$ can be expressed through these ones and therefore also satisfy \eqref{requirement_nulling}. This completes the proof that the requirements \eqref{requirement_cyclic}, \eqref{requirement_cyclic_2} and \eqref{requirement_nulling} are necessary for the construction of a holomorphic vertex $\Upsilon(\omega C^n)$ that vanishes at $y=0$. On the other hand, section \ref{proj-comp_0-form_vertices} proves sufficiency of these requirements.
    \section{Leftover \texorpdfstring{$\delta B_n$}{delta Bn} shift}
    \label{leftover_shift}
We want to find all shifts $\delta B_n$ which do not change first $n-1$ orderings of the vertex $\overset{\circ}{\Upsilon}(\w C^n)$. To do this it is efficient to use formalism developed in section \ref{construction}, where we tried to get an arbitrary vertex $\Upsilon(\w C^n)$. Here, as before, we denote by $\overset{\circ}{\Upsilon}(\w C^n)$ a particular vertex obtained with the particular choice of $\overset{\circ}{B_n}$ field. We need to set $\Upsilon_n^{[k]} = \overset{\circ}{\Upsilon}{}_n^{[k]}$ for all $k<n$. Thus, $F^{[k]} := \overset{\circ}{\Upsilon}{}_n^{[k]} - \Upsilon_n^{[k]} = 0, \quad k<n$. Consequently, by virtue of general solution \eqref{arbitrary_vertex_gen_solution} of equations \eqref{arbitrary_vertex_eq_1}, \eqref{arbitrary_vertex_eq_k}
\begin{multline}
       \delta B_n(y|p_\w; p_1, \dots, p_n) =
       \\
       = \sum_{k=0}^{n-1}  F^{[k]}(0|y + \pi_k; 0, \dots, 0, p_{k+1}, \dots, p_n) e^{-i \left[y\pi_k + \Sigma_k \right]} +  A e^{-i \left[y \pi_n + \Sigma_n \right]} =
    A e^{-i \left[y \pi_n + \Sigma_n \right]},
\end{multline}
where $A$ is an arbitrary constant, we conclude that all shifts $\delta B_n$ that do not change first $n-1$ orderings of $\Upsilon(\w C^n)$ are proportional to $e^{-i \left[y \pi_n + \Sigma_n \right]}$. This conclusion is justified, because for $F^{[k]}=0$, equations \eqref{arbitrary_vertex_eq_1} and \eqref{arbitrary_vertex_eq_k} simply state that $\delta B_n$ does not contribute to $\Upsilon_n^{[k]}$ for any $k<n$.

    \section{Explicit form of \texorpdfstring{$\Upsilon(\w C^3)$}{U(wC3)}}
    \label{explicit_form}
Let us explicitly write down obtained in this paper holomorphic spin-local vertex $\Upsilon(\w C^3)$ using \eqref{local_form_k<n}, \eqref{local_form_n} and results of \cite{1909}.
\begin{multline}
\Upsilon_3^{[0]} =  \frac{\eta^2}{4} \int_{[0,1]^2} d\sigma d\sigma' \, \sigma \sigma' \int_{[0,1]^3} d\tau_1 d\tau_2 d\tau_3 \; \delta\left(1-\sum_k \tau_k \right) \, \left( y p_\w \right)^2 \tau_1 \exp i \left[ - (\tau_2 \sigma + \tau_3 \sigma' + \tau_1 \sigma \sigma') y p_\w - \right. 
\\
\left. -y p_3 (1 - (1 - \tau_3) \sigma) - p_\w p_3 \tau_3 \sigma' 
 + p_1 y \tau_2 \sigma  + p_1 p_\w (1 - \sigma' (1 - \tau_2)) - p_2 y \tau_1 \sigma - p_2p_\w \sigma' \tau_1 \right]
 \label{wCCC}
\end{multline}

\begin{multline}
\Upsilon_3^{[1]} = \frac{\eta^2}{4} \int_{[0,1]^2} \mathrm{d}\sigma \mathrm{d}\sigma' \, \sigma \sigma' \int_{[0,1]^3} d\tau_1 d\tau_2 d\tau_3 \; \delta\left(1-\sum_k \tau_k \right) \, \left( y p_\w \right)^2 \Bigg\{ -\tau_1 \exp i \Big[ (-\tau_2 \sigma + \tau_3 \sigma' + \tau_1 \sigma \sigma') y p_\w  - 
\\
\begin{aligned}
 -y p_3 (1-(1-\tau_3)\sigma ) + p_1 y \tau_2 \sigma- p_1 p_\w (1-\sigma'(1-\tau_2)) + p_\w p_3 \tau_3 \sigma' - p_2  y \tau_1 \sigma + p_2 p_\w \sigma' \tau_1 \Big] & \\
-\tau_1 \exp i \Big[ (\tau_2 \sigma + \tau_3 \sigma' - \tau_1 \sigma \sigma')y p_\w -
\\
- y p_3 (1-(1-\tau_3)\sigma) + p_1 y \tau_1 \sigma  - p_1 p_\w \sigma' \tau_1 + p_\w p_3 \tau_3 \sigma' - p_2 y \tau_2 \sigma  + p_2 p_\w (1-\sigma'(1-\tau_2)) \Big] & \\
-\tau_3 \exp i \Big[ (\tau_2 \sigma - \tau_1 \sigma' + \tau_3 \sigma \sigma') y p_\w -
\\
-y p_3 \tau_3 \sigma - p_1 p_\w \sigma' \tau_1  + p_1 y (1-\sigma(1-\tau_1)) + p_\w p_3 \tau_3 \sigma'- p_2 y \tau_2 \sigma + p_2 p_\w (1-\sigma'(1-\tau_2)) \Big] \Bigg\} &
\end{aligned}
\label{CwCC}
\end{multline}

\begin{multline}
\Upsilon_3^{[2]} = \frac{\eta^2}{4} \int_{[0,1]^2} \mathrm{d}\sigma \mathrm{d}\sigma' \, \sigma \sigma' \int_{[0,1]^3} d\tau_1 d\tau_2 d\tau_3 \; \delta\left(1-\sum_k \tau_k \right) \, \left( y p_\w \right)^2 
\Biggl\{ \tau_1 \exp i \Bigl[ -(-\tau_2 \sigma + \tau_3 \sigma' - \tau_1 \sigma \sigma') y p_\w - 
\\
\begin{aligned}
- y p_3 (1 - (1 - \tau_3) \sigma) 
+ p_1y \tau_1 \sigma + p_1 p_\w \sigma' \tau_1 - p_2 y \tau_2 \sigma - p_2 p_\w (1 - \sigma' (1 - \tau_2)) - p_\w p_3 \tau_3 \sigma' \Bigr] & \\
+ \tau_1 \exp i \Bigl[ (\tau_2 \sigma - \tau_3 \sigma' + \tau_1 \sigma \sigma') y p_\w  - &
\\
- y p_3 \tau_3 \sigma'  
+ p_1 p_\w \tau_2 \sigma + p_1 y (1 - \sigma' (1 - \tau_2)) - p_2 p_\w \tau_1 \sigma - p_2 y \sigma' \tau_1 - p_\w p_3 (1 - (1 - \tau_3) \sigma) \Bigr] \\
+ \tau_3 \exp i \Bigl[ (\tau_2 \sigma + \tau_1 \sigma' - \tau_3 \sigma \sigma') y p_\w - & 
\\
- y p_3 \tau_3 \sigma  
+ p_1 p_\w \tau_1 \sigma' + p_1 y (1 - \sigma (1 - \tau_1)) - p_2 y \tau_2 \sigma - p_2 p_\w (1 - \sigma' (1-\tau_2))- p_\w p_3 \tau_3 \sigma' \Bigr] \Biggr\} &   
\end{aligned}
\label{CCwC}
\end{multline}

\begin{multline}
\Upsilon_3^{[3]} = - \frac{\eta^2}{4} \int_{[0,1]^2} d\sigma d\sigma' \, \sigma \sigma' \int_{[0,1]^3} d\tau_1 d\tau_2 d\tau_3 \; \delta\left(1-\sum_k \tau_k \right) \, \left( y p_\w \right)^2
\tau_1 \exp i \left[ -(\tau_2\sigma + \tau_3\sigma' + \tau_1\sigma\sigma')y p_\w + \right. \\
 p_2 p_\w \tau_1\sigma' - p_2 y \sigma \tau_1 + p_3 y \tau_2\sigma - p_2 p_\w (1 - \sigma'(1 - \tau_2)) + 
\left. p_\w p_1 \tau_3\sigma'- y p_1 (1 - \sigma(1 - \tau_3)) \right]
\label{CCCw}
\end{multline}
\end{appendices}

\end{document}